%% file: main.tex
\documentclass[times,5p,twocolumn,final]{elsarticle}

\usepackage{geometry}
\usepackage{framed,multirow}
\usepackage{amsmath}
\usepackage{booktabs}
\usepackage{longtable,afterpage}
\usepackage{hyperref}
\usepackage{fontawesome5}
\usepackage{graphicx}
\usepackage{multirow}
\usepackage{pbox}
\usepackage{subcaption}
\usepackage[export]{adjustbox}

\usepackage{fancyhdr}
\usepackage{amssymb}
\usepackage{latexsym}

\usepackage{url}
\usepackage{xcolor}

\usepackage{hyperref}

\definecolor{newcolor}{rgb}{.8,.349,.1}

\journal{Medical Image Analysis}
\newcommand{\cxrft}{ChestX-ray14}

\DeclareUnicodeCharacter{2212}{-}

\begin{document}

\begin{frontmatter}

\title{Deep Learning for Chest X-ray Analysis: A Survey}%
\tnotetext[tnote1]{Ecem Sogancioglu and Erdi \c{C}all\i\ contributed equally.}

\author[1]{Ecem Sogancioglu*}
\cortext[cor1]{Corresponding author: 
      E-mail: erdi.calli@radboudumc.nl}
\author[1]{Erdi \c{C}all\i*}
\author[1]{Bram van Ginneken}
\author[1]{Kicky G. van Leeuwen}
\author[1]{Keelin Murphy}

\address[1]{Radboud University Medical Center, Institute for Health Sciences, Department of Medical Imaging, Nijmegen, The Netherlands}

\begin{abstract}
Recent advances in deep learning have led to a promising performance in many medical image analysis tasks. As the most commonly performed radiological exam, chest radiographs are a particularly important modality for which a variety of applications have been researched.  The release of multiple, large, publicly available chest X-ray datasets in recent years has encouraged research interest and boosted the number of publications. In this paper, we review all studies using deep learning on chest radiographs, categorizing works by task: image-level prediction (classification and regression), segmentation, localization, image generation and domain adaptation. Commercially available applications are detailed, and a comprehensive discussion of the current state of the art and potential future directions are provided.
\end{abstract}


\end{frontmatter}

\section{Introduction}
A cornerstone of radiological imaging for many decades, chest radiography (chest X-ray, CXR) remains the most commonly performed radiological exam in the world with industrialized countries reporting an average 238 erect-view chest X-ray images acquired per 1000 of population annually \citep{UniN08}. In 2006, it is estimated that 129 million CXR images were acquired in the United States alone \citep{Mett09}. The demand for, and availability of, CXR images may be attributed to their cost-effectiveness and low radiation dose, combined with a reasonable sensitivity to a wide variety of pathologies. The CXR is often the first imaging study acquired and remains central to screening, diagnosis, and management of a broad range of conditions \citep{Raoo12}. 

\begin{figure*}[h!]
\begin{subfigure}{0.32\textwidth}
    \includegraphics[height=0.9\textwidth,right]{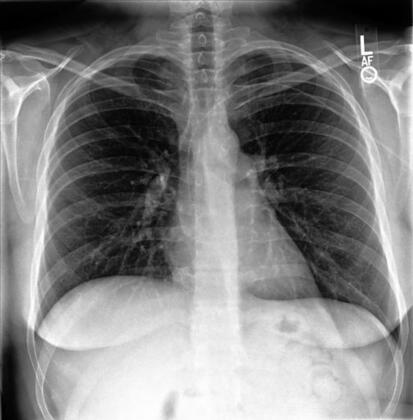}
\end{subfigure}\hfill
\begin{subfigure}{0.32\textwidth}
    \includegraphics[height=0.9\textwidth,center]{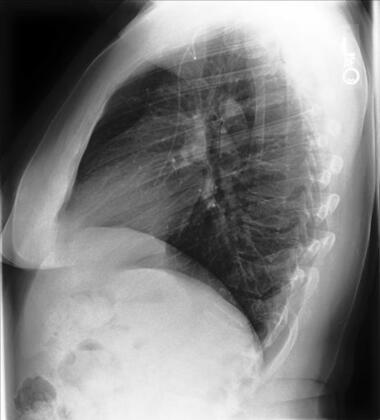}
  
\end{subfigure}\hfill
\begin{subfigure}{0.35\textwidth}
    \includegraphics[height=0.822\textwidth,left]{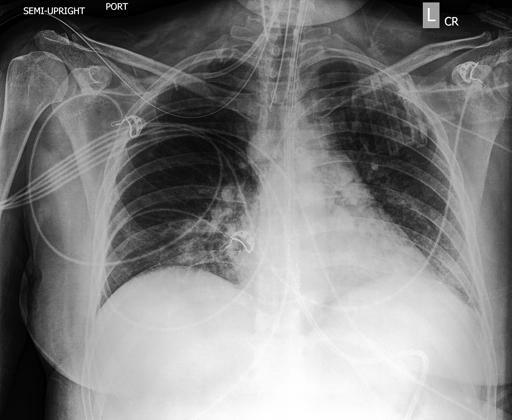}
  
\end{subfigure}\hfill

\caption{Left: posterior-anterior (PA) view frontal chest radiograph. Middle: lateral chest radiograph. Right: Anterior-posterior (AP) view chest radiograph. All three CXRs are taken from the CheXpert dataset \citep{Irvi19}, patient 184. }
\label{fig:xray_example}
\end{figure*}

Chest X-rays may be divided into three principal types, according to the position and orientation of the patient relative to the X-ray source and detector panel: posteroanterior, anteroposterior, lateral. The posteroanterior (PA) and anteroposterior (AP) views are both considered as frontal, with the X-ray source positioned to the rear or front of the patient respectively. The AP image is typically acquired from patients in the supine position, while the patient is usually standing erect for the PA image acquisition. The lateral image is usually acquired in combination with a PA image, and projects the X-ray from one side of the patient to the other, typically from right to left. Examples of these image types are depicted in Figure~\ref{fig:xray_example}.

The interpretation of the chest radiograph can be challenging due to the superimposition of anatomical structures along the projection direction. This effect can make it very difficult to detect abnormalities in particular locations (for example, a nodule posterior to the heart in a frontal CXR), to detect small or subtle abnormalities, or to accurately distinguish between different pathological patterns. For these reasons, radiologists typically show high inter-observer variability in their analysis of CXR images \citep{Quek01,Bala05,Youn94}.

The volume of CXR images acquired, the complexity of their interpretation, and their value in clinical practice have long motivated researchers to build automated algorithms for CXR analysis. Indeed, this has been an area of research interest since the 1960s when the first papers describing an automated abnormality detection system on CXR images were published \citep{Lodw63, Beck64, Meye64, Krug72, Tori73}. The potential gains from automated CXR analysis include increased sensitivity for subtle findings, prioritization of time-sensitive cases, automation of tedious daily tasks, and provision of analysis in situations where radiologists are not available (e.g., the developing world).

In recent years, deep learning has become the technique of choice for image analysis tasks and made a tremendous impact in the field of medical imaging \citep{Litj17}. Deep learning is notoriously data-hungry and the CXR research community has benefited from the publication of numerous large labeled databases in recent years, predominantly enabled by the generation of labels through automatic parsing of radiology reports. This trend began in 2017 with the release of 112,000 images from the NIH clinical center \citep{Wang17a}. In 2019 alone, more than 755,000 images were released in 3 labelled databases (CheXpert \citep{Irvi19}, MIMIC-CXR \citep{John19}, PadChest \citep{Bust20}). In this work, we demonstrate the impact of these data releases on the number of deep learning publications in the field. 

There have been previous reviews on the field of deep learning in medical image analysis \citep{Litj17,Ginn17,Sahi18,Feng19} and on deep learning or computer-aided diagnosis for CXR \citep{Qin18,Kall19,Anis20}. However, recent reviews of deep learning in chest radiography are far from exhaustive in terms of the literature and methodology surveyed, the description of the public datasets available, or the discussion of future potential and trends in the field. The literature review in this work includes 295 papers, published between 2015 and 2021, and categorized by application. A comprehensive list of public datasets is also provided, including numbers and types of images and labels as well as some discussion and caveats regarding various aspects of these datasets. Trends and gaps in the field are described, important contributions discussed, and potential future research directions identified. We additionally discuss the commercial software available for chest radiograph analysis and consider how research efforts can best be translated to the clinic.

The initial selection of literature to be included in this review was obtained as follows: A selection of papers was created using a PubMed search for papers with the following query.
\begin{verbatim}
  chest and ("x-ray" or xray or radiograph) and
  ("deep learning" or cnn or "convolutional" or
  "neural network")
\end{verbatim} 

A systematic search of the titles of conference proceedings from SPIE, MICCAI, ISBI, MIDL and EMBC was also performed, searching paper titles for the same search terms listed above. In the case of multiple publications of the same paper, only the latest publication was included. Relevant peer-reviewed articles suggested by co-authors and colleagues were added. The last search was performed on March 3rd, 2021.

This search strategy resulted in 767 listed papers. Of these, 61 were removed as they were duplicates of others in the list.  A further 261 were excluded as their subject matter did not relate to deep learning for CXR, they were commentary or evaluation papers or they were not written in English.  Publications that were not peer-reviewed were also excluded (8). Finally, during the review process 142 papers were excluded as the scientific content was considered unsound, as detailed further in Section~\ref{sec:discussion}, leaving 295 papers in the final literature review.  

The remainder of this work is structured as follows: Section~\ref{sec:overview_deeplearning} provides a brief introduction to the concept of deep learning and the main network architectures encountered in the current literature. In Section~\ref{sec:datasets}, the public datasets available are described in detail, to provide context for the literature study. The review of the collected literature is provided in Section~\ref{sec:deeplearning_cxr}, categorized according to the major themes identified. Commercial systems available for chest radiograph analysis are described in Section~\ref{sec:commercial}. The paper concludes in Section~\ref{sec:discussion}, with a comprehensive discussion of the current state of the art for deep learning in CXR as well as the potential for future directions in both research and commercial environments.

\section{Overview of Deep Learning Methods}\label{sec:overview_deeplearning}
This section provides an introduction to deep learning for image analysis, and particularly the network architectures most frequently encountered in the literature reviewed in this work. Formal definitions and more in-depth mathematical explanations of fully-connected and convolutional neural-networks are provided in many other works, including a recent review of deep learning in medical image analysis \citep{Litj17}. In this work, we provide only a brief overview of these fundamental details and refer the interested reader to previous literature.

Deep learning is a branch of machine learning, which is a general term describing learning algorithms. The algorithm underpinning all deep learning methods is the neural network, in this case, constructed with many hidden layers (`deep'). These networks may be constructed in many ways with different types of layers included and the overall construction of a network is referred to as its `architecture'.  Sections ~\ref{sec:overview_image_level_prediction} to ~\ref{sec:overview_generative} describe commonly used architectures categorized by types of application in the CXR literature. 

\subsection{Convolutional Neural Networks} 
In the 1980s, networks using convolutional layers were first introduced for image analysis \citep{Fuku82}, and the idea was formalized over the following years \citep{LeCu98}. These convolutional layers now form the basis for all deep learning image analysis tasks, almost without exception. Convolutional layers use neurons that connect only to a small `receptive field' from the previous layer. These neurons are applied to different regions of the previous layer, operating as a sliding window over all regions, and effectively detecting the same local pattern in each location. In this way, spatial information is preserved and the learned weights are shared. 

\subsection{Transfer Learning}
Transfer learning investigates how to transfer knowledge extracted from one domain (source domain) to another (target) domain. One of the most commonly used transfer learning approaches in CXR analysis is the use of pre-training. 

With the pre-training approach, the network architecture is first trained on a large dataset for a different task, and the trained weights are then used as an initialization for the subsequent task for fine-tuning \citep{Yosi14}. Depending on data availability from the target domain, all layers can be re-trained, or only the final (fully connected) layer can be re-trained. This approach allows neural networks to be trained for new tasks using relatively smaller datasets since useful low-level features are learned from the source domain data. It has been shown that pre-training on the ImageNet dataset (for classification of natural images) \citep{Balt19a} is beneficial for chest radiography analysis and this type of transfer learning is prominently used in the research surveyed in this work. ImageNet pre-trained versions of many architectures are publicly available as part of popular deep learning frameworks. The pre-trained architectures may also be used as feature extractors, in combination with more traditional methods, such as support vector machines or random forests. Domain adaptation is another subfield of transfer learning and is discussed thoroughly in Section \ref{sec:overview_domain_adaptation}.

\subsection{Image-level Prediction Networks}\label{sec:overview_image_level_prediction}
In this work we use the term `image-level prediction' to refer to tasks where prediction of a category label (classification) or continuous value (regression) is implemented by analysis of an entire CXR image.  These methods are distinct from those which make predictions regarding small patches or segmented regions of an image. Classification and regression tasks are grouped together in this work since they typically use the same types of architecture, differing only in the final output layer. 
One of the early successful deep convolutional architectures for image-level prediction was AlexNet \citep{Kriz12}, which consists of 5 convolutional layers followed by 3 fully connected layers. AlexNet became extremely influential in the literature when it beat all other competitors in the ILSVRC (ImageNet) challenge \citep{Deng09} by a large margin in 2012. Since then many deep convolutional neural network architectures have been proposed. The VGG family of models \citep{Simo14} use 8 to 19 convolutional layers followed by 3 fully-connected layers. The Inception architecture was first introduced in 2015 \citep{Szeg15} using multiple convolutional filter sizes within layered blocks known as Inception modules. In 2016, the ResNet family of models \citep{He16} began to gain popularity and improve upon previous benchmarks. These models define residual blocks consisting of multiple convolution operations, with skip connections which typically improve model performance. After the success of ResNet, skip connections were widely adopted in many architectures. DenseNet models \citep{Huan17}, introduced in 2017, also use skip connections between blocks, but connect all layers to each other within blocks. A later version of the Inception architecture also added skip connections (Inception-Resnet)  \citep{Szeg16}. The Xception network architecture \citep{Chol17} builds upon the Inception architecture but separates the convolutions performed in the 2D image space from those performed across channels. This was demonstrated to improve performance compared to Inception V3. 

The majority of works surveyed in this review use one or more of the model architectures discussed here with varying numbers of hidden layers.

\subsection{Segmentation Networks}\label{sec:overview_image_segmentation}
Segmentation is a task where pixels are assigned a category label, and can also be considered as a pixel classification. In natural image analysis, this task is often referred to as `semantic segmentation' and frequently requires every pixel in the image to have a specified category. In the medical imaging domain these labels typically correspond to anatomical features (e.g., heart, lungs, ribs), abnormalities (e.g., tumor, opacity) or foreign objects (e.g., tubes, catheters). It is typical in the medical imaging literature to segment just one object of interest, essentially assigning the category `other' to all remaining pixels. 

Early approaches to segmentation using deep learning used standard convolutional architectures designed for classification tasks \citep{chen18b}. These were employed to classify each pixel in a patch using a sliding window approach. The main drawback to this approach is that neighboring patches have huge overlap in pixels, resulting in inefficiency caused by repeating the same convolutions many times. It additionally treats each pixel separately which results in the method being computationally expensive and only applicable to small images or patches from an image. 

To address these drawbacks, fully convolutional networks (FCNs) were proposed, replacing fully connected layers with convolutional layers \citep{Shel17}. This results in a network which can take larger images as input and produces a likelihood map output instead of an output for a single pixel. In 2015, a fully convolutional architecture known as the U-Net was proposed \citep{Ronn15} and this work has become the most cited paper in the history of medical image analysis. The U-Net consists of several convolutional layers in a contracting (downsampling) path, followed by further convolutional layers in an expanding (upsampling) path which restores the result to the input resolution. It additionally uses skip connections between the same levels on the contracting and expanding paths to recover fine details that were lost during the pooling operation. The majority of image segmentation works in this review employ a variant of the FCN or the U-Net.

\subsection{Localization Networks}\label{sec:overview_localization}
This survey uses the term localization to refer to identification of a specific region within the image, typically indicated by a bounding box, or by a point location. As with the segmentation task, localization, in the medical domain, can be used to identify anatomical regions, abnormalities, or foreign object structures. There are relatively few papers in the CXR literature reviewed here that deal specifically with a localization method, however, since it is an important task in medical imaging, and may be easier to achieve than a precise segmentation, we categorize these works together. 

In 2014, the RCNN (Region Convolutional Neural Network) was introduced \citep{Girs14}, identifying regions of interest in the image and using a CNN architecture to extract features of these regions. A support vector machine (SVM) was used to classify the regions based on the extracted features. This method involves several stages and is relatively slow. It was later superseded by fast-RCNN \citep{Girs15} and subsequently by faster-RCNN \citep{Ren17} which streamlined the processing pipeline, removing the need for initial region identification or SVM classification, and improving both speed and performance. In 2017, a further extension was added to faster-RCNN to additionally enable a precise segmentation of the item identified within the bounding box.  This method is referred to as Mask R-CNN \citep{He17}. While this is technically a segmentation network, we mention it here as part of the RCNN family. Another architecture which has been popular in object localization is YOLO (You Only Look Once), first introduced in 2016 \citep{Redm16} as a single-stage object detection method, and improved in subsequent versions in 2017 and 2018 \citep{Redm17, Redm18}. The original YOLO architecture, using a single CNN and an image-grid to specify outputs was significantly faster than its contemporaries but not quite as accurate. The improved versions leveraged both classification and detection training data and introduced a number of training improvements to achieve state of the art performance while remaining faster than its competitors. A final localization network that features in medical imaging literature is RetinaNet \citep{Lin17}. Like YOLO, this is a single stage detector, which introduces the concept of a focal loss function, forcing the network to concentrate on more difficult examples during training. Most of the localization works included in this review use one of the architectures described above.

\subsection{Image Generation Networks}\label{sec:overview_generative}
One of the tasks deep learning has been commonly used for is the generation of new, realistic images, based on information learned from a training set. There are numerous reasons to generate images in the medical domain, including generation of more easily interpretable images (by increasing resolution, or removal of projected structures impeding analysis), generation of new images for training (data augmentation), or conversion of images to emulate appearances from a different domain (domain adaptation). Various generative schemes have also been used to improve the performance of tasks such as abnormality detection and segmentation.

Image generation was first popularized with the introduction of the generative adversarial network (GAN) in 2014 \citep{good14}. The GAN consists of two network architectures, an image generator, and a discriminator which attempts to differentiate generated images from real ones. These two networks are trained in an adversarial scheme, where the generator attempts to fool the discriminator by learning to generate the most realistic images possible while the discriminator reacts by progressively learning an improved differentiation between real and generated images.

The training process for GANs can be unstable with no guarantee of convergence, and numerous researchers have investigated stabilization and improvements of the basic method \citep{Sali16, Heus17, Karr18, Arjo17}. GANs have also been adapted to conditional data generation \citep{Chen16,Oden17} by incorporating class labels, image-to-image translation (conditioned on an image in this case) \citep{Isol17}, and unpaired image-to-image translation (CycleGAN \cite{Zhu17}). 

GANs have received a lot of attention in the medical imaging community and several papers were published for medical image analysis applications in recent years \citep{Yi19}. Many of the image generation works identified in this review employed GAN based architectures.

\subsection{Domain Adaptation Networks}\label{sec:overview_domain_adaptation}
In this work we use the term `Domain Adaptation', which is a subfield of transfer learning, to cover methods attempting to solve the issue that architectures trained on data from a single `domain' typically perform poorly when tested on data from other domains. The term `domain' is weakly defined; In medical imaging it may suggest data from a specific hardware (scanner), set of acquisition parameters, reconstruction method or hospital. It could, less frequently, also refer to characteristics of the population included, for example the gender, ethnicity, age or even strain of some pathology included in the dataset.  

Domain adaptation methods consider a network trained for an image analysis task on data from one domain (the source domain), and how to perform this analysis accurately on a different domain (the target domain). These methods can be categorized as supervised, unsupervised, and semi-supervised depending on the availability of labels from the target domain and they have been investigated for a variety of CXR applications from organ segmentation to multi-label abnormality classification. There is no specific architecture that is typical for domain adaptation, but rather architectures are combined in various ways to achieve the goal of learning to analyze images from unseen domains. The approaches to this problem can be broadly divided into three classes (following the categorization of \citep{Wang18a}); discrepancy-based, reconstruction-based and adversarial-based. 

Discrepancy-based approaches aim to induce alignment between the source and target domain in some feature space by fine-tuning the image analysis network and optimizing a measurement of discrepancy between the two domains. Reconstruction-based approaches, on the other hand, use an auxiliary encoder-decoder reconstruction network that aims to learn domain invariant representation through a shared encoder. Adversarial-based approaches are based on the concept of adversarial training from GANs, and use a discriminator network which tries to distinguish between samples from the source and target domains, to encourage the use of domain-invariant features. This category of approaches is the most commonly used in CXR analysis for domain adaptation, and consists of generative and non-generative models. Generative models transform source images to resemble target images by operating directly on pixel space whereas non-generative models use the labels on the source domain and leverage adversarial training to obtain domain invariant representations.

\section{Datasets}\label{sec:datasets}
Deep learning relies on large amounts of annotated data. 
The digitization of radiological workflows enables medical institutions to collate and categorize large sets of digital images. In addition, advances in natural language processing (NLP) algorithms mean that radiological reports can now be automatically analyzed to extract labels of interest for each image. These factors have enabled the construction and release of multiple large labelled CXR datasets in recent years. Other labelling strategies have included the attachment of the entire radiology report and/or labels generated in other ways, such as radiological review of the image, radiological review of the report, or laboratory test results. Some datasets include segmentations of specified structures or localization information.

In this section we detail each public dataset that is encountered in the literature included in this review as well as any others available to the best of our knowledge. Details are provided in Table~\ref{tab:datasets}. Each dataset is given an acronym which is used in the literature review tables (Tables ~\ref{tab:image_level_predictions_table} to ~\ref{tab:other_table}) to indicate that the dataset was used in the specified work.

\input{datasets_table}

\begin{enumerate}

    \item \cxrft\ (C) is a dataset consisting of $112,120$ CXRs from $30,805$ patients \citep{Wang17a}. The CXRs are collected at the (US) National Institute of Health. The images are distributed as 8-bit grayscale images scaled to $1024 \times 1024$ pixels. The dataset was automatically labeled from radiology reports, indicating the existence of 14 types of abnormality. 
    
    \item CheXpert (X) is a dataset consisting of $224,316$ CXRs from $65,240$ patients \citep{Irvi19}. The CXRs are collected at Stanford Hospital between October 2002 and July 2017. The images are distributed as 8-bit grayscale images with original resolution. The dataset was automatically labeled from radiology reports using a rule-based labeler, indicating the presence, absence, uncertainty, and no-mention of 12 abnormalities, no findings, and the existence of support devices.
        
    \item MIMIC-CXR (M) is a dataset consisting of $371,920$ CXRs from and $64,588$ patients \citep{John19}. The CXRs are collected from patients admitted to the emergency department of Beth Israel Deaconess Medical Center between 2011 and 2016. In version 1 (V1) the images are distributed as 8-bit grayscale images in full resolution. The dataset was automatically labeled from radiology reports using the same rule-based labeler system (described above) as CheXpert. A second version (V2) of MIMIC-CXR was later released including the anonymized radiology reports and DICOM files. 
    
    \item PadChest (P) is a dataset consisting of $160,868$ CXRs from $109,931$ studies and $67,000$ patients \citep{Bust20}. The CXRs are collected at San Juan Hospital (Spain) from 2009 to 2017. The images are stored as 16-bit grayscale images with full resolution. $27,593$ of the reports were manually labeled by physicians. Using these labels, an RNN was trained and used to label the rest of the dataset from the reports. The reports were used to extract 174  findings, 19 diagnoses, and 104 anatomic locations. The labels conform to a hierarchical taxonomy based on the standard Unified Medical Language System (UMLS) \citep{Bode04}. 
    
     \item PLCO (PL) is a screening trial for prostate, lung, colorectal and ovarian (PLCO) cancer \citep{Zhu13}. The lung arm of this study has $185,421$ CXRs from $56,071$ patients. The NIH distributes a standard set of $25,000$ patients and $88,847$ frontal CXRs. This dataset contains 22 disease labels with 4 abnormality levels and the locations of the abnormalities. 
     
    \item Open-i (O) is a dataset consisting of $7,910$ CXRs from $3,955$ studies and $3,955$ patients \citep{Demn12}. The CXRs are collected from the Indiana Network for Patient Care \citep{Mcdo05}. The images are distributed as anonymized DICOMs. The radiological findings obtained by radiologist interpretation are available in MeSH format\footnote{\url{https://www.nlm.nih.gov/mesh/meshhome.html}}.
    
    \item Ped-Pneumonia (PP) is a dataset consisting of 5,856 pediatric CXRs \citep{Kerm18}. The CXRs are collected from Guangzhou Women and Children’s Medical Center, Guangzhou, China. The images are distributed in 8-bit grayscale images scaled in various resolutions. The labels include bacterial and viral pneumonia as well as normal.
    
    \item JSRT dataset (J) consists of 247 images with a resolution of $2048 \times 2048$, 0.175mm pixel-size and 12-bit depth \citep{Shir00}. It includes nodule locations (on 154 images) and diagnosis (malignant or benign). The reference standard for heart and lung segmentations of these images are provided by the SCR dataset \citep{Ginn06a} and we group these datasets together in this work. 
    
    \item RSNA-Pneumonia (RP) is a dataset consisting of $30,000$ CXRs with pneumonia annotations \citep{RSNA18}. These images are acquired from \cxrft\ and are 8-bit grayscale with $1024 \times 1024$ resolution. Annotations are added by radiologists using bounding boxes around lung opacities and 3 classes indicating normal, lung opacity, not normal.
    
    \item Shenzhen (S) is a dataset consisting of 662 CXRs \citep{Jaeg14}. The CXRs are collected at Shenzhen No.3 Hospital in Shenzhen, Guangdong providence, China in September 2012. The images, including some pediatric images, are distributed as 8-bit grayscale with full resolution and are annotated for signs of tuberculosis.
    
    \item Montgomery (MO) is a dataset consisting of 138 CXRs \citep{Jaeg14}. The CXRs are collected by the tuberculosis control program of the Department of Health and Human Services of Montgomery County, MD, USA. The images are distributed as anonymized DICOMs, annotated for signs of tuberculosis and additionally include lung segmentation masks.

    \item BIMCV (B) is a COVID-19 dataset released by the Valencian Region Medical ImageBank (BIMCV) in 2020 \citep{Vaya20}. It includes CXR images as well as CT scans and laboratory test results. The dataset includes 3,293 CXRs from 1,305 COVID-19 positive subjects. CXR images are 16-bit PNG format with original resolution. 
    
    \item COVIDDSL (CD) is a COVID-19 dataset released by the HM Hospitales group in Spain \citep{CDSL20}. It includes CXR images for 1,725 patients as well as detailed results from laboratory testing, vital signs etc. All subjects are stated to be confirmed COVID-19 positive. 
    
    \item COVIDGR (CG) is a dataset consisting of 852 PA CXR images where half of them are labeled as COVID-19 positive based on corresponding RT-PCR results obtained within at most 24 hours \citep{Tabi20}. This dataset was collected from Hospital Universitario Clínico San Cecilio, Granada, Spain, and the level of severity of positive cases is provided.
        
    \item SIIM-ACR (SI) This dataset was released for a Kaggle challenge on pneumothorax detection and segmentation \citep{ACR19}.  Researchers have determined that at least some (possibly all) of the images are from the \cxrft\ dataset although the challenge organizers have not confirmed the data sources. They are supplied in $1024 \times 1024$ resolution as DICOM files. Pixel segmentations of the pneumothorax in positive cases are provided.
    
    \item CXR14-Rad-Labels (CR) supplies additional annotations for a subset of \cxrft\ data \citep{Majk20}. It consists of 4 labels for 4,374 studies and 1,709 patients. These labels are collected by the adjudicated agreement of 3 radiologists. These radiologists were selected from a cohort of 11 radiologists for the validation split (2,412 studies from 835 patients), and 13 radiologists for the test split (1,962 studies from 860 patients). The individual labels from each radiologist as well as the agreement labels were provided.
    
    \item COVID-CXR (CV) is a dataset consisting of 930 CXRs at the time of writing (the dataset remains in continuous development) \citep{Cohe20b}. The CXRs are collected from a large variety of locations using different methods including screenshots from papers researching COVID-19. Available labels vary accordingly, depending on what information is available from the source where the image was obtained. Images do not have a standard resolution and are published as 8-bit PNG or JPEG files.

    \item NLST (N) is a dataset of publicly available CXRs collected during the NLST screening trial \cite{Nati11}. This trial aimed to compare the use of low-dose computed tomography (CT) with CXRs for lung cancer screening in smokers. The study had 26,732 participants in the CXR arm and a part of this data is available upon request. 
    
    \item Object-CXR (OB) is a dataset of 10,000 CXR images from hospitals in China with foreign objects annotated on the images.  The download location (https://jfhealthcare.github.io/object-CXR/) is no longer available at the time of writing. Further detail is not provided since it cannot be verified from the image source. 
      
    \item Belarus (BL) This dataset is included since it is used in a number of reviewed papers however the download location (http://tuberculosis.by) is no longer available at the time of writing. The dataset consisted of approximately 300 frontal chest X-rays with confirmed TB. Further detail is not provided since it can no longer be verified from the image source. 
\end{enumerate}

\begin{figure}[h!]
    \centering
    \includegraphics[width=0.5\textwidth]{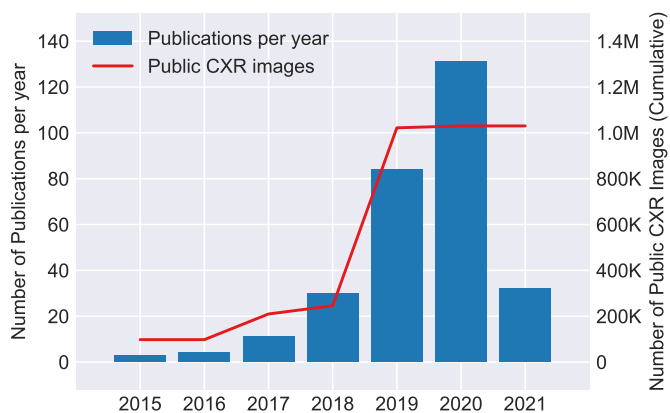}
    \caption{Number of publications that were reviewed in this work, by year, compared with the number of publicly available CXR images. Data for 2021 is until March 3rd of that year.}
    \label{fig:studies_and_images_per_year}
\end{figure}

The rapid increase in the number of publicly available CXR images in recent years has positively impacted the number of deep learning studies published in the field. Figure \ref{fig:studies_and_images_per_year} illustrates the cumulative number of publicly available CXR images and the number of publications on deep learning with CXR per year.

\subsection{Public Dataset Caution}
Publication of medical image data is extremely important for the research community in terms of advancing the state of the art in deep learning applications.  However, there are a number of caveats that should be considered and understood when using the public datasets described in this work.  Firstly, many datasets make use of Natural Language Processing (NLP) to create labels for each image.  Although this is a fast and inexpensive method of labeling, it is well known that there are inaccuracies in labels acquired this way \citep{Irvi19, Oakd20, Oakd19}.  There are a number of causes for such inaccuracies. Firstly, some visible abnormalities may not be mentioned in the radiology report, depending on the context in which it was acquired \citep{Olat19}. Further, the NLP algorithm can be erroneous in itself, interpreting negative statements as positive, failing to identify acronyms, etc. Finally, many findings on CXR are subtle or doubtful, leading to disagreements even among expert observers \citep{Olat19}. Acknowledging some of these issues, \cite{Irvi19} includes labels for uncertainty or no-mention in the labels on the CheXpert dataset.  One particular cause for concern with NLP labels is the issue of systematic or structured mislabeling, where an abnormality is consistently labeled incorrectly in the same way. An example of this occurs in the ChestX-ray14 dataset where subcutaneous emphysema is frequently identified as (pulmonary) `emphysema' \citep{Call19, Oakd20}. 

It has been demonstrated that deep neural networks can tolerate reasonable levels of label inaccuracy in the training set without a significant effect on model performance \citep{Call19, Roln18}.  Although such labels can be used for training, for an accurate evaluation and comparison of models it is desirable that the test dataset is accurately labelled. In the literature reviewed in this work, many authors rely on labels from NLP algorithms in their test data, while others use radiologist annotations, laboratory tests and/or CT verification for improved test set labelling. We refer to data that uses these improved labelling techniques as gold standard data (Table \ref{tab:datasets}).

The labels defined in the public datasets should also be considered carefully and understood by the researchers using them.  Many labels have substantial dependencies between them. For example, some datasets supply labels for both `consolidation' and `pneumonia'. Consolidation (blocked airspace) is an indicator of a patient with pneumonia, suggesting there will be significant overlap between these labels.  A further point for consideration is that, in practice, not all labels can be predicted by a CXR image alone.  Pneumonia is rarely diagnosed by imaging alone, requiring other clinical signs or symptoms to suggest that this is the cause for a visible consolidation. 

Many public datasets release images with a lower quality than is used for radiological reading in the clinic.  This may be a cause for decreased performance in deep learning systems, particularly for more subtle abnormalities.  The reduction in quality is usually related to a decrease in image size or bit-depth prior to release.  This is typically carried out to decrease the overall download size of a dataset. However, in some cases, CXR data has been collected by acquiring screenshots from online literature, which results in an unquantifiable degradation of the data.  In the clinical workflow, DICOM files are the industry standard for storing CXRs, typically using 12 bits per pixel and with image dimensions of approximately 2 to 4 thousand pixels in each of the X and Y directions.  In the event that the data is post-processed before release it would be desirable that a precise description of all steps is provided to enable researchers to reproduce them for dataset combination.  

\section{Deep Learning for Chest Radiography}\label{sec:deeplearning_cxr}
\begin{figure}[h!]
    \centering
    \includegraphics[width=0.5\textwidth]{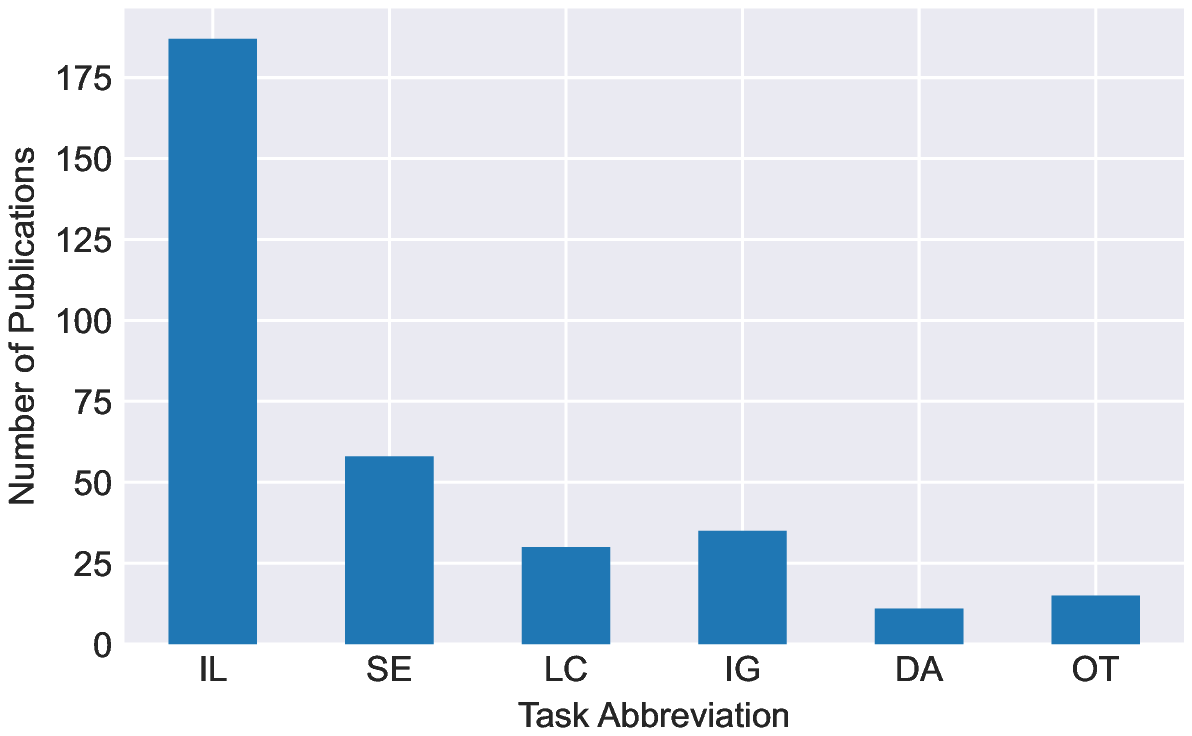}
    \caption{Number of publications reviewed for each task. 295 studies are included, each study may perform at most two tasks.\\ \textbf{Tasks}: IL=Image-level Predictions, SE=Segmentation, LC=Localization, IG=Image Generation, DA=Domain Adaptation, OT=Other.}
    \label{fig:publications_per_task}
\end{figure}

In this section we survey the literature on deep learning for chest radiography, dividing it into sections according to the type of task that is addressed (Image-level Prediction, Segmentation, Image Generation, Domain Adaptation, Localization, Other). For each of these sections a table detailing the literature on that task is provided. Some works which have equal main focus on two tasks may appear in both tables.  For Segmentation and Localization, only studies that quantitatively evaluate their results are included in those categories. Figure~\ref{fig:publications_per_task} shows the number of studies for each of the tasks.

\input{subsections/image_level_prediction_studies}

\subsection{Segmentation}\label{sec:studies_seg}

\begin{figure}[h!]
    \centering
    \includegraphics[scale=0.7]{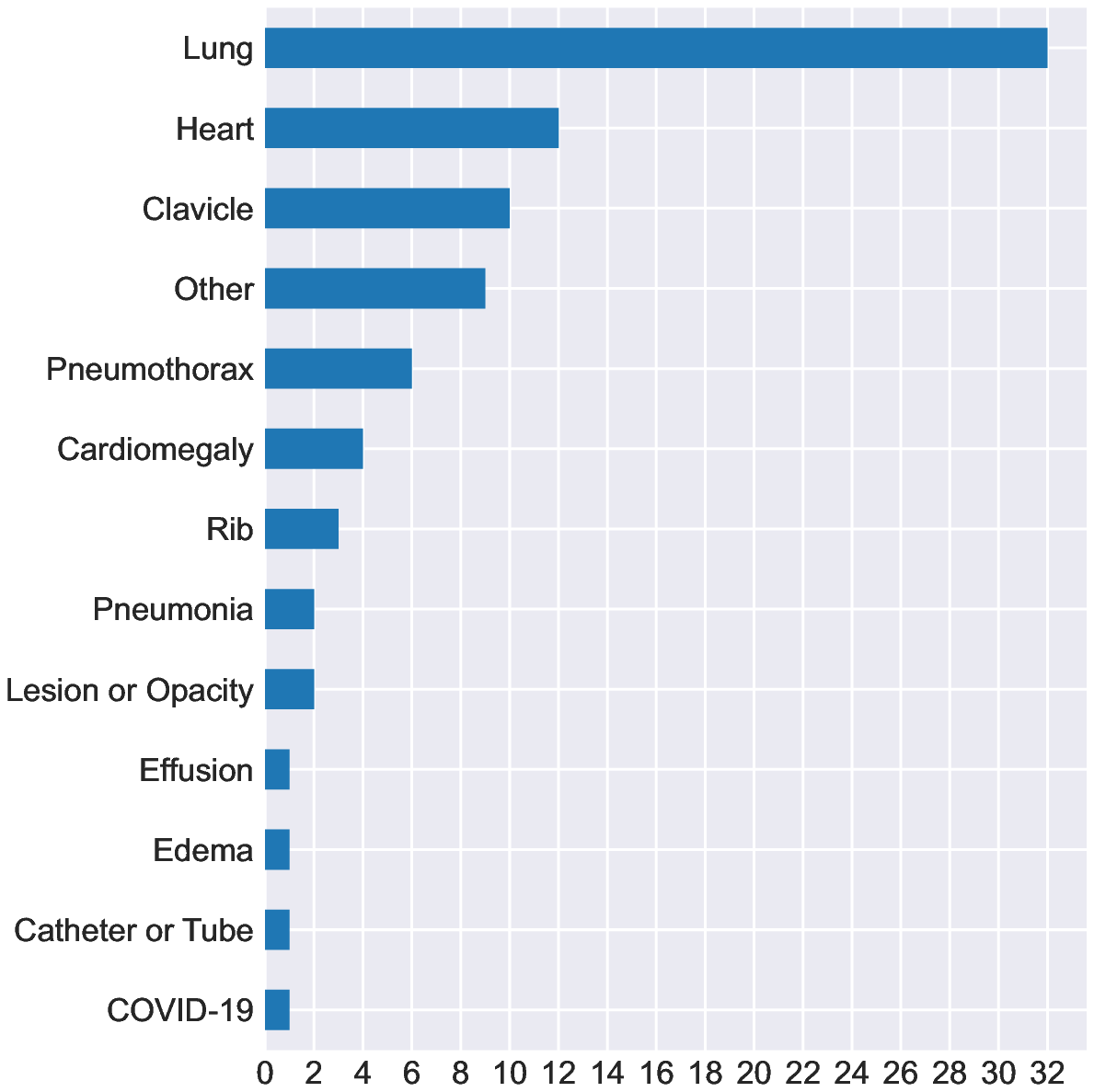}
    \caption{Number of studies for the Segmentation labels. 58 papers are included, each may study more than one label.}
    \label{fig:segmentation/disease_counts}
\end{figure}

Segmentation is one of the most commonly studied subjects in CXR analysis (58 papers) and includes literature focused on the identification of anatomy, foreign objects or abnormalities. The segmentation literature reviewed for this work is detailed fully in Table~\ref{tab:segmentation_table}. Anatomical segmentation of the heart, lungs, clavicles or ribs, on chest radiographs, is a core part of many computer aided detection (CAD) pipelines. It is typically used as an initial step of such pipelines to define the region of interest for subsequent image analysis tasks to improve performance and efficiency \citep{Balt19a,Wang20a,Raja19a, Heo19, Liu19, Mans16}. Further, the segmentation itself can be useful to quantify clinical parameters based on shape or area measurements. For example, cardiothoracic ratio, a clinically used measurement to assess heart enlargement (cardiomegaly), can be directly calculated from heart and lung segmentations \citep{Soga20,Li19}. Organ segmentation has, for these reasons, become one of the most commonly studied subjects among CXR segmentation tasks as seen in Figure~\ref{fig:segmentation/disease_counts}. 

\input{tables/Segmentation}

Another application found in the CXR literature is foreign object segmentation, i.e. catheter, tubes, lines, for which high performance levels have been reported using deep learning \citep{Lee18, Frid19, Sull20}. 

Interestingly, only a small number of works addressed segmentation of abnormalities. \cite{Hurt20} focused on segmentation of pneumonia, and \cite{Tolk20} developed a method to segment pneumothorax. Both of these works used recently published challenge datasets (hosted by Kaggle), namely RSNA-Pneumonia and SIIM-ACR. In general, the determination of abnormal locations on CXR is dominated by methods which addressed this as a localization task (i.e. via bounding-box type annotations) rather than exact delineation of abnormalities through segmentation. This is likely to be attributable to the difficulty of precise annotation on a projection image and to the high annotation cost for precise segmentations. 

A small number of works tackled the segmentation task using a patch-based CNN, which is trained to classify the center of pixel in the patch as foreground or background by means of sliding-window approach \citep{Arba17, Souz19}. However, this approach is generally considered inefficient for segmentation and most works use fully convolutional networks (FCN) \citep{Shel17}, which can take larger, arbitrary sized, images as input and produce a similar sized, per-pixel prediction, likelihood map in a single forward pass. In particular, the U-Net architecture \citep{Ronn15}, a type of FCN, dominates the field with 50\% of segmentation works in literature (29/58) employing it or some similar variant. Successful applications were built with this architecture to segment organs \citep{Novi18,Furu19,Kita19}, pneumonia \citep{Hurt20} and foreign objects \citep{Sull20, Frid19}. For example, \cite{Novi18} compared three U-Net variant architectures for multi-class segmentation of the heart, clavicles and lungs on the JSRT dataset. Using regularization to prevent over-fitting and weighted cross entropy loss to balance the dataset, they outperformed the human observer at heart and lung segmentation. This result was in line with other works \citep{Khol20, Bort19, Arsa20} employing FCN-type architectures which also achieved very high performance levels on this dataset.

One commonly encountered challenge is that many algorithms produce noisy segmentation maps. In order to tackle this, several works employed post-processing techniques. \cite{Lee18} used a probabilistic Hough line transform algorithm to remove false positives and produce a smoother segmentation of peripherally inserted central catheters (PICC). \cite{Groz20} used a heuristic approach to average cross-fold predictions with an optimized binarization threshold and a dilation technique for pneumothorax segmentation. Some authors proposed to learn post-processing by training an independent network, inputting segmentation predictions for refinement, rather than using conventional methods. For example, \cite{Larr20} used denoising autoencoders, trained to produce anatomically plausible segmentations from the initial predictions. Similarly, \cite{Souz19} used a FCN to refine segmentation predictions. The final segmentation was achieved by combining the initial and reconstructed segmentation results.

A number of researchers used a multi-stage training strategy, where network predictions are refined in several steps during training \citep{Wess19, Souz19, Xue18, Xue20}. For example, \cite{Xue18} employed faster-RCNN to produce coarse segmentation results, which were then used to crop the images to a region of interest, which was provided to a U-Net trained to predict the final segmentation result. Similarly, \cite{Souz19} employed two networks, where the second network received the predictions of the first to refine the segmentation results. \cite{Wess19} trained separate networks for segmentation of each rib in chest radiographs based on Mask R-CNN. The predicted segmentation results from the rib above was fed to each network as an additional input.

Although most of the works in the literature harnessed FCN architectures, a few authors employed recurrent neural networks (RNN) for segmentation tasks \citep{Yi20a, Mill18, Math19} and report good performance. \cite{Mill18} proposed a novel architecture where the decoding component was long short term memory (LSTM) architecture to obtain multi-scale feature integration. The proposed approach achieved a Dice score of 0.97 for lung segmentation on Montgomery dataset. Similarly, \cite{Yi19} developed a scale RNN, a network based on encoder and decoder architecture with recurrent modules, for segmentation of catheter and tubes on pediatric chest X-rays.

The high cost of obtaining segmentation annotations motivates the development of segmentation systems which incorporate weak-labels or simulated datasets with the aim of reducing annotation costs \citep{Frid19,Ouya19,Lu20, Yi20a}. Several works addressed this using weakly supervised learning approaches \citep{Lu20,Ouya19}. \cite{Lu20} proposed a graph convolutional network based architecture which required only one labeled image and leveraged large amounts of unlabeled data (one-shot learning) through a newly introduced three contour-based loss function. \cite{Ouya19} proposed a pneumothorax segmentation framework which incorporated both images with pixel level annotations and weak image-level annotations. The authors trained an image classification network, ResNet-101, with weakly labeled data to derive attention maps. These attention maps were then used to train a segmentation model, Tiramisu, together with pixel level annotations.

\subsection{Localization}\label{sec:studies_loc}

Localization refers to the identification of a region of interest using a bounding box or point coordinates rather than a more specific pixel segmentation. In this section we discuss only the CXR localization literature which provides a quantitative evaluation of this task. It should be noted that there are many other works which train networks for an image-level prediction task and provide some examples of heatmaps (e.g., saliency map or GradCAM) to suggest which region of the image determines the label. While this may be considered as a form of localization, these heatmaps are rarely quantitatively evaluated and such works are not included here. Table~\ref{tab:localization_table} details all the reviewed studies where localization was a primary focus of the work.

\input{tables/Localization}

\begin{figure}[h!]
    \centering
    \includegraphics[scale=0.7]{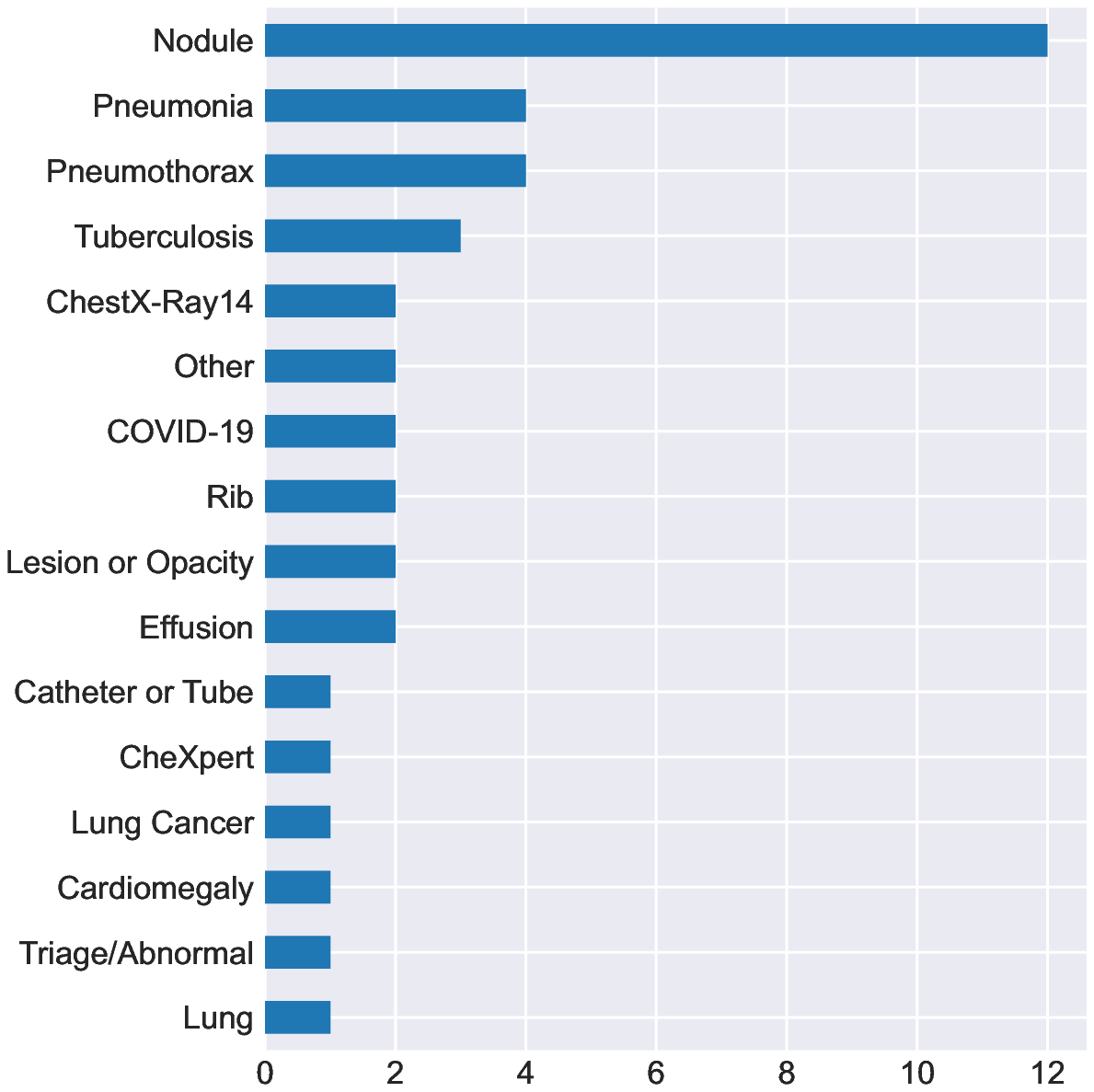}
    \caption{Number of studies for the Localization labels. 30 papers are included, each may study more than one label.}
    \label{fig:localization/disease_counts}
\end{figure}

The majority of CXR analysis papers performing localization focus on identifying abnormalities rather than objects (e.g., catheter) or anatomy (e.g., ribs). Localization of nodules, tuberculosis and pneumonia are commonly studied applications in the literature, as illustrated in Figure~\ref{fig:localization/disease_counts}. 

In recent years, a variety of specific architectures, i.e. YOLO, Mask R-CNN, Faster R-CNN, have been designed in computer vision research aiming at developing more accurate and faster algorithms for localization tasks \citep{Zhao19}. Such state of the art architectures  have been rapidly adapted for CXR analysis and shown to achieve high-level performance. For example, \cite{Park19} demonstrated that the (original) YOLO architecture was successful at identifying the location of pneumothorax on chest radiographs. The model was evaluated on an external dataset with CXRs from 1,319 patients which were obtained after percutaneous transthoracic needle biopsy (PTNB) for pulmonary lesions; it achieved an AUC of 0.898 and 0.905 on 3-h and 1-day follow-up chest radiographs, respectively. Similarly, other studies \citep{Kim20, Schu20, Take19, Kim19} harnessed architectures like RetinaNet, Mask R-CNN and RCNN for localization of nodules and masses. \cite{Kim20} trained RetinaNet and Mask R-CNN for detection of nodule and mass and investigated the optimal input size. The authors showed that, using a square image with 896 pixels as the edge length, RetinaNet and Mask R-CNN achieved FROC of 0.906 and 0.869, respectively.

A number of papers adapted classification architectures (e.g., ResNet, DenseNet) to directly regress landmark locations for CXR localization tasks \citep{Hwan19,Cha19a}. One common way of tackling this is to adapt the networks to produce heatmap predictions and draw boxes around the areas that created the highest signals. For example, \cite{Hwan19} tailored a DenseNet-based classifier to produce heatmap predictions for each of four types of CXR abnormalities. The network was trained with pixel-wise cross entropy between the predictions and annotations. Similarly, \cite{Cha19a} adapted ResNet-50 and ResNet-101 architectures for localization of nodules and masses on CXR. Other studies \citep{Xue18a, Li20} tackled this problem using patch-based approaches, commonly referred as multiple instance learning, creating patches from chest X-rays and evaluating these for the presence of abnormalities. 

One challenge in building robust deep learning localization systems is to collect large annotated datasets. Collecting such annotations is time-consuming and costly which has motivated researchers to build systems incorporating weaker labels during training. This research area is referred to as weakly supervised learning, and has been investigated by numerous works \citep{Hwan16a,Hwan19,Nam19,Pesc19, Tagh19} for localization of a variety of abnormalities in CXR. Most of the works \citep{Hwan19,Pesc19, Nam19, Hwan16a} leveraged weak image-level labels by adapting a CNN architecture to create two branches for localization (heatmap predictions) and classification. A hybrid loss function was used, combining localization and classification losses, which enabled training of the networks using images without localization annotations.

\subsection{Image Generation}\label{sec:studies_gen}
There are 35 studies identified in this work whose main focus is Image Generation, as detailed in Table~\ref{tab:image_generation_table}. Image generation techniques have been harnessed for a wide variety of purposes including data augmentation \citep{Sale19}, visualization \citep{Bigo19,Seah19}, abnormality detection through reconstruction \citep{Tang19,Woll20}, domain adaptation \citep{Zhan18} or image enhancement techniques \citep{Lee19}. 

\input{tables/Image-Generation}

The generative adversarial network (GAN) \citep{good14,Yi19} has became the method of choice for image generation in CXR and over 50\% of the works reviewed here used GAN-based models.

A number of works focused on CXR generation to augment training datasets \citep{Mora18, Zhan19, Sale19} by using unconditional GANs which synthesize images from random noise. For example, \cite{Sale19} trained a DCGAN model, similar to \cite{Mora18}, independently for each class, to generate chest radiographs with five different abnormalities. The authors demonstrated that this augmentation process improved the abnormality classification performance of DCNN classifiers (ResNet, GoogleNet, AlexNet) by balancing the dataset classes. Another work \citep{Zhan19} proposed a novel GAN architecture to improve the quality of generated CXR by forcing the generator to learn different image representations. The authors proposed SkrGAN, where a sketch prior constraint is introduced by decomposing the generator into two modules for generating a sketched structural representation and the CXR image, respectively.

Abnormality detection is another task which has been addressed through a combination of image generation and one-class learning methods \citep{Tang19,Mao20}. The underlying idea of these methods is that a generative model trained to reconstruct healthy images will have a high reconstruction error if abnormal images are input at test time, allowing them to be identified. \cite{Tang19} harnessed GANs and employed a U-Net type autoencoder to reconstruct images (as the generator), and a CNN-based discriminator and encoder. The discriminator received both reconstructed images and real images to provide supervisory signal for realistic reconstruction through adversarial training.

Similarly, \cite{Mao20} proposed an autoencoder for abnormality detection which was trained only with healthy images. In this case the autoencoder was tailored to not only reconstruct healthy images but also produce uncertainty predictions. By leveraging uncertainty, the authors proposed a normalized reconstruction error to distinguish abnormal CXR images from normal ones.

The most widely studied subject in the image generation literature is image enhancement. Several researchers investigated bone suppression \citep{Liu20, Mats20a, Zars19,Goze20,Lin20,Zhou20a} and lung enhancement \citep{Li19a,Goze20} techniques to improve image interpretability. A number of works \citep{Liu20, Zhou20a} employed GANs to generate bone-suppressed images. For example, \cite{Liu20} employed GANs and leveraged additional input to the generator to guide the dual-energy subtraction (DES) soft-tissue image generation process. In this study, bones, edges and clavicles were first segmented by a CNN model, and the resulting edge maps were fed to the generator with the original CXR image as prior knowledge. For building a deep learning model for bone suppressed CXR generation, the paired dual energy (DE) imaging is needed, which is not always available in abundance. Several other studies \citep{Li19a, Goze20} addressed this by leveraging digitally reconstructed radiographs for enhancing the lungs and bones in CXR. For instance, \cite{Li19a} trained an autoencoder for generating CXR with bone suppression and lung enhancement, and the knowledge obtained from DRR images were integrated through the encoder.

\subsection{Domain Adaptation}\label{sec:studies_da}

Most of the papers surveyed in this work train and test their method on data from the same domain. This finding is inline with the previously reported studies \citep{Kim19,Prev19} and highlights an important concern: most of the performance levels reported in the literature might not generalize well to data from other domains \citep{Zech18}. Several studies \citep{Yao19,Zech18,Cohe20c} demonstrated that there was a significant drop in performance when deep learning systems were tested on datasets outside their training domain for a variety of CXR applications. For example, \cite{Yao19} investigated the performance of a DenseNet model for abnormality classification on CXR images using 10 diverse datasets varied by their location and patient distributions. The authors empirically demonstrated that there was a substantial drop in performance when a model was trained on a single dataset and tested on the other domains. \cite{Zech18} observed a similar finding for pneumonia detection on chest radiographs. 

Domain adaptation (DA) methods investigate how to improve the performance of a model on a dataset from a different domain than the training set. In CXR analysis, DA methods have been investigated in three main settings; adaptation of CXR images acquired from different hardware, adaptation of pediatric to adult CXR and adaptation of digitally reconstructed radiographs (generated by average intensity projections from CT) to real CXR images. All domain adaptation studies, and studies on generalization reviewed in this work are detailed in Table~\ref{tab:domain_adaptation_table}.

\input{tables/Domain-Adaptation}

Most of the research on DA for CXR analysis harnessed adversarial-based DA methods, which either use generative models (e.g., CycleGANs) or non-generative models to adapt to new domains using a variety of different approaches. For example, \cite{Dong18} investigated an unsupervised domain adaptation based on adversarial training for lung and heart segmentation. In this approach, a discriminator network, ResNet, learned to discriminate between segmentation predictions (heart and lung) from the target domain and reference standard segmentations from the source domain. This approach forced the FCN-based segmentation network to learn domain invariant features and produce realistic segmentation maps. A number of works \citep{Chen18a, Zhan18,Oliv18} addressed unsupervised DA using CycleGAN-based models to transform source images to resemble those from the target domain. For example, \cite{Zhan18} used a CycleGAN-based architecture to adapt CXR images to digitally reconstructed radiographs (DRR) (generated from CT scans), for anatomy segmentation in CXR. A CycleGAN-based model was employed to convert the CXR image appearance and a U-Net variant architecture to simultaneously segment organs of interest. Similarly, CycleGAN-based models were adapted to transfer DRR images to resemble CXR images for bone segmentation \citep{Oliv20a} and to transform adult CXR to pediatric CXR for pneumonia classification \citep{Tang19}.  

Unlike most of the studies which utilized DA methods in unsupervised setting, a few studies considered supervised and semi-supervised approaches to adapt to the target domain. \cite{Oliv20b} employed a MUNIT-based architecture \citep{Huan18} to map target images to resemble source images, subsequently feeding the transformed images to the segmentation model. The authors investigated both unsupervised and semi-supervised approaches in this work, where some labels from the target domain were available. Another work by \cite{Leng20} studied several recently proposed continual learning approaches, namely joint training, elastic weight consolidation and learning without forgetting, to improve the performance on a target domain and to mitigate effectively catastrophic forgetting for the source domain. The authors evaluated these methods for 2 publicly available datasets, \cxrft\ and MIMIC-CXR, for a multi-class abnormality classification task and demonstrated that joint training achieved the best performance.

\subsection{Other Applications}\label{sec:studies_oth}

In this section we review articles with a primary application that does not fit into any of the categories detailed in Sections \ref{sec:studies_ilp} to \ref{sec:studies_da} (14 studies). These works are detailed fully in Table~\ref{tab:other_table}. 

\input{tables/Other}

Image retrieval is a task investigated by a number of authors \citep{Anav15,Anav16,Conj17,Chen18,Silv20,Owai20,Haq21}. The aim of image retrieval tools is to search an image archive to find cases similar to a particular index image. Such algorithms are envisaged as a tool for radiologists in their daily workflow. \cite{Chen18} proposed a ranked feature extraction and hashing model, while \cite{Silv20} proposed to use saliency maps as a similarity measure.

Another task that did not belong to previously defined categories is out-of-distribution detection. Studies working on this \citep{Marq19,Call19a,Bozo20} aim to verify whether a test sample belongs to the distribution of the training dataset as model performance is otherwise expected to be sub-optimal. \cite{Call19a} propose using the training dataset statistics on different layers of a deep learning model and applying Mahalanobis distance to see the distance of a sample from the training dataset. \cite{Bozo20} approach the problem differently and train an unsupervised autoencoder. Later they use the feature encodings extracted from CXRs to define a database of known encodings and compare new samples to this database.    

Report generation is another task which has attracted interest in deep learning for CXR \citep{Li20c,Yuan19,Syed20,Xue18b}. These studies aim to partially automate the radiology workflow by evaluating the chest X-ray and producing a text radiology report. For example, \cite{Syed20} first determines the findings to be reported and then makes use of a large dataset of existing reports to find a similar case. This case report is then customized to produce the final output. 

One other task of interest is image registration \citep{Mans20}. This task aims to find the geometric transformation to convert a CXR so that it anatomically aligns with another CXR image or a statistically defined shape. The clinical goal of this task is typically to illustrate interval change between two images. Detecting new findings, tracking the course of a disease, or evaluating the efficacy of a treatment are among the many uses of image registration \citep{Vier16}. To that end, \cite{Mans20} aims to create an anatomically plausible registration by using the heart and lung segmentations to guide the registration process.

\section{Commercial Products}\label{sec:commercial}

Computer-aided analysis of CXR images has been researched for many years, and in fact CXR was one of the first modalities for which a commercial product for automatic analyis became available in 2008.  In spite of this promising start, and of the advances in the field achieved by deep learning, translation to clinical practice, even as an assistant to the reader, is relatively slow. There are a variety of legal and ethical considerations which may partly account for this \citep{Recht20, Stro20}, however there is growing acceptance that artificial intelligence (AI) products have a place in the radiological workflow and attempts are underway to understand and address the issues to be overcome \citep{Chok19}. In this section we examine the currently available commercial products for CXR analysis. 

An up to date list of commercial products for medical image analysis \citep{GCPR20, Leeu21} was searched for products applicable to chest X-ray. One product was excluded as it is not specifically a CXR diagnostic tool, but a texture analysis product for many modalities. The 21 remaining products are listed in Table~\ref{tab:commercial}. A number of these products have already been evaluated in peer-reviewed publications, as shown in Table~\ref{tab:commercial} and it is beyond the scope of this work to make an assessment of their performance.  All of the listed products are CE marked (Europe) and/or FDA cleared (United States) and are thus available for clinical use \citep{GCPR20,Leeu21}.

\input{tables/commercial_table}

The commercial products include applications for a wide range of abnormalities, with 6 of them reporting results for more than 5 (and up to 30) different labels. The most commonly addressed task is pneumothorax identification (8 products), followed by pleural effusion (7), nodules (6) and tuberculosis (4). In contrast with the literature, which is dominated by image-level prediction algorithms, 17 of 21 products in Table~\ref{tab:commercial} claim to provide localization of one or more abnormalities which they are designed to detect, usually visualized with heatmaps or contouring of abnormalities. Two further products are designed for generation of bone suppression images, one for interval change visualization and one for identification and reporting of healthy images. Products contribute differently to the workflow of the radiologist. Five products focus on detecting acute cases to prioritize the worklist and speed up time to diagnosis. Draft reports are produced by five other products, for either the normal (healthy) cases only or for all cases.  The production of draft reports, like workflow prioritization, is aimed at optimizing the speed and efficiency of the radiologist.

\section{Discussion}\label{sec:discussion}

In this work we have detailed datasets, literature and commercial products relevant to deep learning in CXR analysis. It is clear that this area of research has thrived on the release of multiple large, public, labeled datasets in recent years, with 209 of 295 publications reviewed here using one or more public datasets in their research. The number of publications in the field has grown consistently as more public data becomes available, as demonstrated in Figure~\ref{fig:studies_and_images_per_year}. However, although these datasets are extremely valuable, there are multiple caveats to be considered in relation to their use, as described in Section~\ref{sec:datasets}.  In particular, the caution required in the use of NLP-extracted labels is often overlooked by researchers, especially for the evaluation and comparison of models.  For accurate assessment of model performance, the use of `gold-standard' test data labels is recommended.  These labels can be acquired through expert radiological interpretation of CXRs (preferably with multiple readers) or via associated CT scans, laboratory test results, or other appropriate measurements.  

Other important factors to be considered when using public data include the image quality (if it has been reduced prior to release, is this a limiting factor for the application?) and the potential overlap between labels.  Although a few publications address label dependencies, this is most often overlooked, frequently resulting in the loss of valuable diagnostic information.

While the increased interest in CXR analysis following the release of public datasets is a positive development in the field, a secondary consequence of this readily available labeled data is the appearance of many publications from researchers with limited experience or understanding of deep learning or CXR analysis.  The literature reviewed during the preparation for this paper was very variable in quality.  A substantial number of the papers included offer limited novel contributions although they are technically sound. Many of these studies report experiments predicting the labels on public datasets using off-the-shelf architectures and without regard to the label inaccuracies and overlap, or the clinical utility of such generic image-level algorithms. A large number of works were excluded for reasons of poor scientific quality (142). In 112 of these the construction of the dataset gave cause for concern, the most common example being that the training dataset was constructed such that images with certain labels came from different data sources, meaning that the images could be easily differentiated by factors other than the label of interest.  In particular, a large number of papers (61) combined adult COVID-19 subjects with pediatric (healthy and other-pneumonia) subjects in an attempt to classify COVID-19.  Other reasons for exclusion included the presentation of results optimized on a validation set (without a held-out test set), or the inclusion of the same images multiple times in the dataset prior to splitting train and test sets.  This latter issue has been exacerbated by the publication of several COVID-19 related datasets which combine data from multiple public sources in one location, and are then themselves combined by authors building deep-learning systems. Such concerns about dataset construction for COVID-19 studies have been discussed in several other works \citep{Lope21,DeGr20,Cruz21,Magu20,Tart20}.

Although a broad range of off-the-shelf architectures are employed in the literature surveyed for this review, there is little evidence to suggest that one architecture outperforms another for any specific task. Many papers evaluate multiple different architectures for their task but differences between the various architecture results are typically small, proper hyperparameter optimization is not usually performed and statistical significance or data-selection influence are rarely considered. Many such evaluations use inaccurate NLP-extracted labels for evaluation which serves to muddy the waters even further. 

While it is not possible to suggest an optimal architecture for a specific task, it is observed that ensembles of networks typically perform better than individual models \citep{Diet00}. At the time of writing, most of the top-10 submissions from the public challenges (CheXpert \citep{Irvi19}, SIIM-ACR \citep{ACR19}, and RSNA-Pneumonia \citep{RSNA18}) consist of network ensembles. There is also promise in the development of self-adapting frameworks such as the nnU-Net \citep{Isen21} which has achieved an excellent performance in many medical image segmentation challenges. This framework adapts specifically to the task at hand by selecting the optimal choice for a number of steps such as preprocessing, hyperparameter optimization, architecture etc., and it is likely that a similar optimization framework would perform well for classification or localization tasks, including those for CXR images.

In spite of the pervasiveness of CXR in clinics worldwide, translation of AI systems for clinical use has been relatively slow. Apart from legal and ethical considerations regarding the use of AI in medical decision making \citep{Recht20, Stro20}, a discussion which is outside the scope of this work, there are still a number of technical hurdles where progress can be made towards the goal of clinical translation.  Firstly, the generalizability of AI algorithms is an important issue which needs further work. A large majority of papers in this review draw training, validation and test samples from the same dataset. However, it is well known that such models tend to have a weaker performance on datasets from external domains. If access to reliable data from multiple domains remains problematic then domain adaptation or active learning methods could be considered to address the generalization issue. An alternative method to utilize data from multiple hospitals without breaching regulatory and privacy codes is federated learning, whereby an algorithm can be trained using data from multiple remote locations \citep{Shel19}. Further research is required to determine how this type of system will work in clinical practice. 

A final issue for deep learning researchers to consider is frequently referred to as `explainable AI'. Systems which produce classification labels without any indication of reasoning raise concerns of trustworthiness for radiologists. It is also significantly faster for experts to accept or reject the findings of an AI system if there is some indication of how the finding was reached (e.g., identification of nodule location with a bounding box, identification of cardiac and thoracic diameters for cardiomegaly detection). Every commercial product for detection of abnormality in CXR provides a localization feature to indicate the abnormal location, however the literature is heavily focused on image-level predictions with relatively few publications where localization is evaluated.  

Beyond the resolution of technical issues, researchers aiming to produce clinically useful systems need to consider the workflow and requirements of the end-user, the radiologist or clinician, more carefully. At present, in the industrialized world, it is expected that an AI system will act, at least initially, as an assistant to (not a replacement for) a radiologist.  As a 2D image, the CXR is already relatively quickly interpreted by a radiologist, and so the challenge for AI researchers is to produce systems that will save the radiologist time, prioritize urgent cases or improve the sensitivity/specificity of their findings.  Image-level classification for a long list of (somewhat arbitrarily defined) labels is unlikely to be clinically useful. Reviewing such a list of labels and associated probabilities for every CXR would require substantial time and effort, without a proportional improvement in diagnostic accuracy. A simple system with bounding boxes indicating abnormal regions is likely to be more helpful in directing the attention of the radiologist and has the potential to increase sensitivity to subtle findings or in difficult regions with many projected structures. Similarly, a system to quickly identify normal cases has the potential to speed up the workflow as identified by multiple vendors and in the literature~\citep{Dyer21,Dunn19, Balt20}. 

To further understand how AI could assist with CXR interpretation, we first must consider the current typical workflow of the radiologist, which notably involves a number of additional inputs beyond the CXR image, that are rarely considered in the research literature. In most scenarios (excluding bedside/AP imaging) both a frontal and lateral CXR are acquired as part of standard imaging protocol, to reduce the interpretation difficulties associated with projected anatomy. Very few studies included in this review made use of the lateral image, although there are indications that it can improve classification accuracy \citep{Hash20}.  Furthermore, the reviewing radiologist has access to the clinical question being asked, the patient history and symptoms and in many cases other supporting data from blood tests or other investigations.  All of this information assists the radiologist to not only identify the visible abnormalities on CXR (e.g., consolidation), but to infer likely causes of these abnormalities (e.g., pneumonia). Incorporation of data from multiple sources along with the CXR image information will almost certainly improve sensitivity and specificity and avoid an algorithm erroneously suggesting labels which are not compatible with data from external sources. Another extremely important and time-consuming element in the radiological review of CXR is comparison with previous images from the same patient, to assess changes over time. Interval change is a topic studied by very few authors and addressed by only a single commercial vendor (by provision of a subtraction image). Innovative AI systems for the visualization and quantification of interval change with one or more previous images could substantially improve the efficiency of the radiologist.  Finally, the radiologist is required to produce a report as a result of the CXR review, which is another time-consuming process addressed by very few researchers and just a handful of commercial vendors. A system which can convert radiological findings to a preliminary report has the potential to save time and cost for the care provider.

In many areas of the world, medical facilities that do perform CXR imaging do not have access to radiological expertise. This presents a further opportunity for AI to play a role in diagnostic pathways, as an assistant to the clinician who is not trained in the interpretation of CXR. Researchers and commercial vendors have already identified the need for AI systems to detect signs of tuberculosis (TB), a condition which is endemic in many parts of the world, and frequently in low-resource settings where radiologists are not available. While such regions of the world could potentially benefit from AI systems to detect other conditions, it is important to identify in advance what conditions could be feasibly both detected and treated in these areas where resources are severely limited.

The findings of this work suggest that while the deep learning community has benefited from large numbers of publicly available CXR images, the direction of the research has been largely determined by the available data and labels, rather than the needs of the clinician or radiologist. Future work, in data provision and labelling, and in deep learning, should have a more direct focus on the clinical needs for AI in CXR interpretation. More accurate comparison and benchmarking of algorithms would be enabled by additional public challenges using appropriately annotated data for clinically relevant tasks.

\section{Acknowledgements}
This work was supported by the Dutch Technology Foundation STW, which formed the NWO Domain Applied and Engineering Sciences and partly funded by the Ministry of Economic Affairs (Perspectief programme P15-26 ‘DLMedIA: Deep Learning for Medical Image Analysis’.

\bibliographystyle{model2-names.bst}\biboptions{authoryear}

\bibliography{bibliography}

\end{document}

%% file: datasets_table.tex
\newcommand{\ReportParsing}{RP}																		
\newcommand{\RadiologistInterpretationReports}{RIR}																		
\newcommand{\RadiologistInterpretation}{RI}																		
\newcommand{\RadiologistCohort}{RCI}																		
\newcommand{\LaboratoryTest}{LT}																		
\newcommand{\BoundingBox}{BB}																		
\newcommand{\Classification}{CL}																		
\newcommand{\ClassificationLocation}{CLoc}																		
\newcommand{\Report}{R}																		
\newcommand{\Segmentation}{SE}																		
																		
\begin{table*}[h!]																		
\centering		
\scriptsize
\caption{CXR datasets available for research. Values above $10,000$ are rounded and shortened using K, indicating thousand (such as 10K for $10,000$). \\ \textbf{Labeling Methods}: \ReportParsing =Report Parsing, \RadiologistInterpretationReports =Radiologist Interpretation of Reports, \RadiologistInterpretation =Radiologist Interpretation of Chest X-Rays, \RadiologistCohort =Radiologist Cohort  agreement on Chest X-Rays, \LaboratoryTest =Laboratory Tests.
\\ \textbf{Annotation Types}: \BoundingBox  =Bounding Box, \Classification =Classification, \ClassificationLocation =Classification with Location label, \Report =Report, \Segmentation =Segmentation. \\ \textbf{Gold Standard Data}: This refers to the number of images labeled by methods other than Report Parsing \\ }			\label{tab:datasets}
\resizebox{\textwidth}{!}{\begin{tabular}{p{3.5cm}llllllll}																		
\toprule																		
	&	\multirow{3}{2.3cm}{\textbf{Patients (P) \\ Studies (S) \\  Images (I)} }	&	\multirow{3}{2cm}{\textbf{View Positions}}	&	\multicolumn{3}{c}{\textbf{Annotation}}			&			\multirow{3}{2cm}{\textbf{Image Format}}	&	\multirow{3}{2cm}{\textbf{Labeling method}}	&	\multirow{3}{2cm}{\textbf{Gold Standard Data}}	\\ \cmidrule(lr){4-6}

	&		&		&	\textbf{Types}	&	\textbf{Labels}	&	\textbf{Studies}	&		&		&		\\	
	&		&		&		&		&		&		&		&		\\ \cmidrule(lr){2-9}

\textbf{\cxrft~(C)}	&	P: 31K	&	PA: 67K	&	\Classification	&	14	&	112K	&	PNG	&	\ReportParsing	&		\\	
 \cite{Wang17a}	&	I: 112K	&	AP: 45K	&	\BoundingBox	&	8	&	983	&		&	\RadiologistInterpretation	&	984	\\	
	&		&		&		&		&		&		&		&		\\ \cmidrule(lr){2-9}

\textbf{CheXpert (X)}	&	P: 65K	&	PA: 29K	&	\Classification	&	14	&	224K	&	JPEG	&	\RadiologistCohort	&	235	\\	
\cite{Irvi19}	&	S: 188K	&	AP: 162K	&		&		&		&		&	\ReportParsing	&		\\	
	&	I: 224K	&	LL: 32K	&		&		&		&		&		&		\\
\cmidrule(lr){2-9}

\textbf{MIMIC-CXR (M)}	&	P: 65K	&	PA+AP: 250K	&	\Classification \ (V1)	&	14	&	372K	&	JPEG (V1)	&	\ReportParsing	&		\\	
\cite{John19}	&	S: 224K	&	LL: 122K	&	\Report \ (V2)	&		&	372K	&	DICOM (V2)	&		&		\\	
	&	I: 372K	&		&		&		&		&		&		&		\\ 
	\cmidrule(lr){2-9}

\textbf{PadChest (P)}	&	P: 67K	&	PA: 96K	&	\Classification	&	193	&	110K	&	DICOM	&	\RadiologistInterpretationReports	&	27593	\\	
\cite{Bust20}	&	S: 110K	&	AP: 20K	&	\Report	&		&	110K	&		&	\ReportParsing	&		\\	
	&	I: 160K	&	LL: 51K	&		&		&		&		&		&		\\ \cmidrule(lr){2-9}

\textbf{PLCO (PL)}	&	P: 25K	&	PA: 89K	&	\Classification	&	22	&	89K	&	TIFF	&	\RadiologistInterpretation	&	All	\\	
\cite{Zhu13}	&	I: 89K	&		&	\ClassificationLocation	&	17	&	89K	&		&		&		\\ \cmidrule(lr){2-9}	

\textbf{Open-i (O)}	 &	P: 3,955	&	PA: 3,955	&	\Report	&		&	3,955	&	DICOM	&	\RadiologistInterpretation	&	All	\\	
\cite{Demn12}	&	I: 7,910	&	LL: 3,955	&		&		&		&		&		&		\\ \cmidrule(lr){2-9}

\textbf{Ped-pneumonia (PP)}	&	I: 5,856	&		&	\Classification	&	2	&	5,856	&	JPEG	&	\RadiologistInterpretation	&	All	\\	
\cite{Kerm18} 	&	&	&		&		&		&		&		&		\\ \cmidrule(lr){2-9}

\textbf{JSRT+SCR (J)}	&	I: 247	&	PA: 247	&	\Segmentation	&	3	&	247	&	DICOM	&	\RadiologistInterpretation	&	All \\	
\cite{Shir00}	&		&		&		&		&		&		&		&		\\ \cmidrule(lr){2-9}

\textbf{RSNA-Pneumonia (RP)}	&	I: 30K	&	PA: 16K	&	\BoundingBox	&	1	&	30K	&	DICOM	&	\RadiologistInterpretation	&	All	\\	
\cite{RSNA18}	&		&	AP:14K	&	\Classification	&		&		&		&		&		\\ \cmidrule(lr){2-9}

\textbf{Shenzhen (S)}	&	I: 340	&	PA: 340	&	\Classification	&	2	&	340	&	DICOM	&	\RadiologistInterpretation	&	All	\\	
\cite{Jaeg14}	&		&		&		&		&	340	&		&		&		\\ \cmidrule(lr){2-9}

\textbf{Montgomery (MO)}	&	I: 138	&	PA: 138	&	\Classification	&	2	&	138	&	PNG	&	\RadiologistInterpretation	&	All	\\	
\cite{Jaeg14}	&		&		&	\Segmentation	&		&	138	&		&		&		\\ \cmidrule(lr){2-9}

\textbf{BIMCV (B)}	&	P: 1,305	&	PA: 1,132 	&	CL	&	1	&	2,391	&	DICOM	&	LT	&	All	\\	
\cite{Vaya20}	&	S: 2,391	&	AP: 1,346	&		&		&		&		&		&		\\ 
                &	I: 3,293    &	LL: 815	    &		&		&		&		&		&		\\ \cmidrule(lr){2-9}

\textbf{COVIDDSL (CD)}	&	P: 1,725	&	PA	&	\Classification	&	1	& 4,943		&	DICOM	& \LaboratoryTest &	All	\\	
\cite{CDSL20}	&	S: 4,943	&	AP (most)	&		&		&		&		&		&		\\ 	
	&		&	LL	&		&		&		&		&		&		\\ \cmidrule(lr){2-9}

\textbf{COVIDGR (CG)}	&	I: 852	&	PA:852	&	\Classification	&	2	&	852	&	JPEG	&	\RadiologistInterpretation	&	All	\\	
\cite{Tabi20} 	&	&	&		&		&		&		&		&		\\ \cmidrule(lr){2-9}

\textbf{SIIM-ACR (SI)}	& I: 16K		&	PA: 11K	&	\Segmentation	&	1	&	16K	&	DICOM	&	\RadiologistInterpretation	&	All	\\	
\cite{ACR19}	&	P:16K	&	AP:	4,799 &		&		&		&		&		&		\\ \cmidrule(lr){2-9}

\textbf{CXR14-Rad-Labels (CR)}		&	P: 1,709	&	AP: 3,244	&	\Classification	&	4	&	4,374	&	PNG	&	\RadiologistCohort	&	All	\\	
\cite{Majk20}&	I: 4,374	&	PA: 1,132	&		&		&		&		&		&		\\ \cmidrule(lr){2-9}

\textbf{COVID-CXR (CC)}	& I:866	&	PA:344	&	\Classification	&		&		&	PNG+JPEG	&	Various	&		\\	
\cite{Cohe20b}	&	P:449	&	AP:438	& \BoundingBox		&		&		&		&		&		\\ 	
	&		&	LL:84	& \Segmentation		&		&		&		&		&		\\ \cmidrule(lr){2-9}

\textbf{NLST (N)}	&	I: 5493	&	\multicolumn{7}{l}{No public information available} 			\\	
\cite{Nati11}	&		&	\multicolumn{7}{l}{Number of images is reported by \cite{Lu19}.}	\\
\cmidrule(lr){2-9}

\textbf{Object-CXR (OB)}	&	I:10K	&	\multicolumn{7}{l}{No longer at original download location} 	\\	
\cmidrule(lr){2-9}	


\textbf{Belarus (BL)}	&	I:~300	&	\multicolumn{7}{l}{No longer at original download location} 	\\

\bottomrule
\end{tabular}}	
\end{table*}																		


%% file: subsections/image_level_prediction_studies.tex
\subsection{Image-level Prediction}\label{sec:studies_ilp}

Image-level prediction refers to the task of predicting a label (classification) or a continuous value (regression) by analyzing an entire image. Classification labels may relate to pathology (e.g. pneumonia, emphysema), information such as the subject gender, or orientation of the image.  Regression values might, for example, indicate a severity score for a particular pathology, or other information such as the age of the subject. 

\begin{figure}[h!]
    \centering
    \includegraphics[scale=0.7]{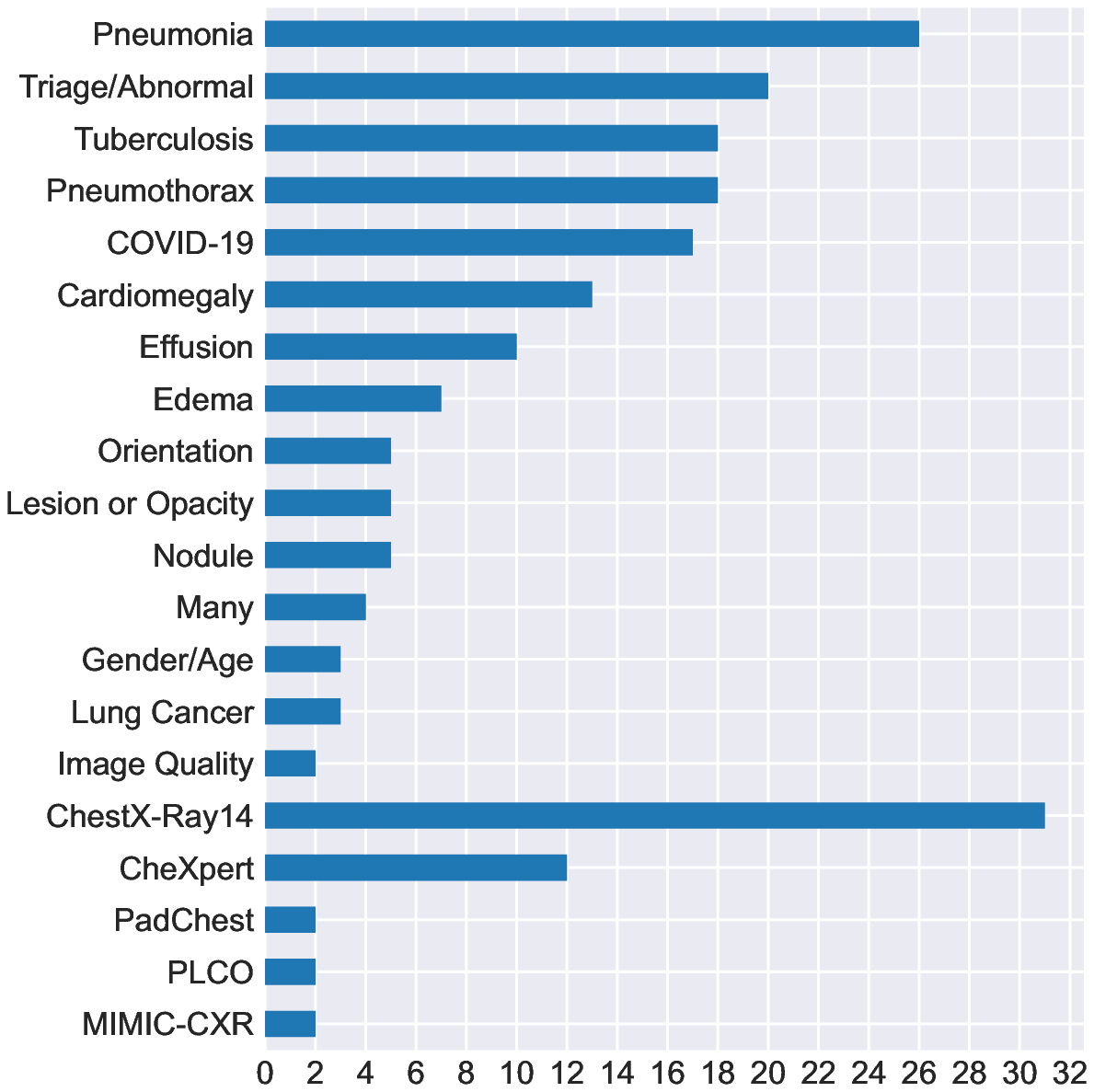}
    \caption{Number of studies for the Image-level Prediction labels. The studies that specifically work on a dataset and its labels are grouped together at the bottom. 187 papers are included, each may study more than one label.}
    \label{fig:classification/disease_counts}
\end{figure}

We classified 187 studies, fully detailed in Table~\ref{tab:image_level_predictions_table} as image-level predictions. Most of these studies make use of off-the shelf deep learning models to predict a pathology, metadata information or a set of labels provided with a dataset. The number of studies for each label are provided in Figure~\ref{fig:classification/disease_counts}.
 
\input{tables/Image-level-Predictions}

The most commonly studied image-level prediction task is predicting the labels of the \cxrft\ dataset (31 studies). For example, \cite{Balt19} compares the performance of various approaches to classify the 14 disease labels provided by the \cxrft\ dataset. \cite{Rajp18} compares the performance of an ensemble of deep learning models to board-certified and resident radiologists, showing that their models achieve a performance comparable to expert observers in most of the 14 labels provided by \cxrft. Following this, pneumonia is  second most studied subject (26 studies). Of the 26 studies that worked with pneumonia, 12 studied pediatric chest X-rays and 11 of those used the Ped-Pneumonia dataset for training and evaluation \citep{Raja18,Yue20,Lian20a,Behz20,Elsh20,Uret20,Mitt20,Shah20,Qu20,Ferr20,Anan20}.
Classification to Normal/Abnormal (or Triage) is another commonly studied topic (20 studies).  Here, studies aim to distinguish normal CXRs or prioritize urgent/critical cases with the goal of reducing the radiologist workload or improving the reporting time. For example, \cite{Anna19} develops a triaging pipeline based on the urgency of exams. Similarly, \cite{Tang20} compares the performance of various deep learning models applied to several public chest X-ray datasets for distinguishing abnormal cases.
Pneumothorax is another commonly studied condition (18 studies). For example, \cite{Tayl18} aims to detect potentially critical patients and proposes that such models can be used to alert clinicians.  Another common topic is tuberculosis detection (18 studies). The first studies that use deep learning to detect this infectious disease are \cite{Hwan16, Lakh17a}. Performance of a deep learning model and how the assistance of this model improves the radiologist performance is studied by \cite{Rajp20}. This study in particular evaluates the use of extra clinical information such as age, white blood cell count, patient temperature and oxygen saturation to assist the deep learning model. 
Diagnosis or evaluation of COVID-19 from CXR is another topic that has attracted a lot of interest from researchers (17 studies). For example, \cite{Cohe20} predicts the disease severity, similarly \cite{Li20b} predicts the disease progression by comparing an exam with the previous exams of the patient, and \cite{Tart20} detects COVID-19 using a very limited amount of data. Other than these most common tasks, there are many studies using deep learning to make Image-level Predictions from CXRs. Other commonly utilized labels are illustrated in Figure~\ref{fig:classification/disease_counts} and listed in Table~\ref{tab:image_level_predictions_table}.

A large proportion of the studies use pre-trained standard architectures that can easily be found in deep learning libraries such as Tensorflow or Pytorch. These architectures are commonly Resnet~\citep{He16}, DenseNet~\citep{Huan17}, Inception~\citep{Szeg15}, VGG~\citep{Simo14}, or AlexNet~\citep{Kriz12}. The choice of model depth (such as ResNet-18, ResNet-50, DenseNet-121, DenseNet-161) also varies between studies as there is no standard in this design choice. Most of those studies do not introduce methodological novelty but report or compare the performances of multiple architectures on a given task. For example, \cite{Desh20} compares various Resnet and Densenet models using both pretrained and randomly initialized weights on the performance of detecting the existence of foreign objects. Similarly many other studies compare the performance of different architectures with various depths on a given task, for example \cite{Sing19,Tang20,Raja20a,Chak20}. Just like the model depth and architecture, there are many factors that affect the performance of a deep learning model. The effect of various data augmentation and input pre-processing methods are evaluated by \cite{Sira19, Vidy19}. The effect of increasing or decreasing the image size are evaluated in \cite{DSou19, Balt19}. Various pre-training schemes are evaluated by \citep{Goze19, Balt19}. More sophisticated pre-processing steps to improve model performance include bone suppression \citep{Balt19a, Zhou20a} and lung cropping \citep{Liu19}. 

Some studies bring methodological novelty by making use of methods that are known to work well to improve model performance elsewhere. For example, it is known that an ensemble of many models improves performance compared to a single model \citep{Diet00}. Some studies that make use of this method are \cite{Rajp18, Raja19, Raja20, Zhan21c}. Attention mining (or object-region mining, attention-based) models are also found in the literature \citep{Wei18}. Those models aim to improve performance and add localization capabilities to an image-level prediction model. Some studies making use of attention mining models are \cite{Cai18, Saed20}. Multiple-instance learning (multi-instance learning or MIL) \citep{Xu14} is another method that is used to add localization capabilities to image-level prediction models. MIL breaks the input image into smaller parts (instances), makes individual predictions relating to those instances and combines this information to make a prediction for the whole image. Some studies that make use of MIL are \citep{Cros20b, Schw20}. Other topics within the literature include model uncertainty \citep{UlA20, Ghes19}, quality of the CXR \citep{Mora20, McMa20,Mora19,Taka20,McMa20} and defence against adversarial attack \citep{Li20a, Anan20, Xue19}.  


The different properties of datasets are also utilized to improve model capabilities or performance. Many of the public datasets make use of labels that are not mutually exclusive.  This has resulted in a number of papers addressing the dependencies among abnormality labels \citep{Pham20, Chen20, Chak20}. Since many of the labels are common between datasets from different institutes there has been investigation of the issues related to domain and/or label shift in images from different sources \citep{Luo20, Cohe20c}. The effect of dataset sizes is evaluated by \cite{Dunn19}. Semi-supervised learning methods combine a small set of labeled and a large set of unlabeled data to train a model \citep{Gyaw19, Gyaw20, Wang19b, Unni20}. 

Most of the studies working on image-level prediction tasks deal with frontal CXR images. The importance of lateral chest X-rays and models that can deal with multiple views are evaluated in \cite{Bert19a, Hash20, Baya20}.


%% file: tables/Image-level-Predictions.tex
\begin{table*}[h!]
\centering
\scriptsize
\caption{Image-Level Prediction Studies (Section \ref{sec:studies_ilp}).\\ 
\textbf{Tasks}: AA=Adversarial Attack, DA=Domain Adaptation, IC=Interval Change, IG=Image Generation, IR=Image Retrieval, LC=Localization, OT=Other, PR=Preprocessing, RP=Report Parsing, SE=Segmentation, WS=Weak Supervision. \textbf{Bold font} in tasks implies that this additional task is central to the work and the study also appears in another table in this paper.\\ 
\textbf{Labels}: C=ChestX-Ray14, CM=Cardiomegaly, CV=COVID, E=Edema, GA=Gender/Age, L=Lung, LC=Lung Cancer, LO=Lesion or Opacity, M=MIMIC-CXR, MN=Many, ND=Nodule, OR=Orientation, P=PadChest, PE=Effusion, PL=PLCO, PM=Pneumonia, PT=Pneumothorax, Q=Image Quality, T=Triage/Abnormal, TB=Tuberculosis, TU=Catheter or Tube, X=CheXpert, Z=Other. \\ 
\textbf{Datasets}: BL=Belarus, C=ChestX-ray14, CC=COVID-CXR, CG=COVIDGR, J=JSRT+SCR, M=MIMIC-CXR, MO=Montgomery, O=Open-i, P=PadChest, PL=PLCO, PP=Ped-pneumonia, PR=Private, RP=RSNA-Pneumonia, S=Shenzen, SI=SIIM-ACR, SM=Simulated CXR from CT, X=CheXpert.}
\label{tab:image_level_predictions_table}
\begin{tabular}{p{3.6cm}p{8.75cm}p{1.2cm}p{1.35cm}p{1.35cm}}
\toprule
      Citation &                                                                                Method &       Other Tasks &       Labels &     Datasets \\
\midrule
  \cite{Liu19} &          Combines lung cropped CXR model and a CXR model to improve model performance & \textbf{SE},LC,PR &          C,L &          C,J \\
 \cite{Soga20} &         Comparison of image-level prediction and segmentation models for cardiomegaly &       \textbf{SE} &           CM &            C \\
 \cite{Que18a} &                  A network with DenseNet and U-Net for classification of cardiomegaly &       \textbf{SE} &           CM &            C \\
   \cite{Li19} &            U-Net based model for heart and lung segmentation for cardiothoracic ratio &       \textbf{SE} &           CM &           PR \\
 \cite{Mora20} &        Combines lung cropped CXR model and a CXR model using the segmentation quality &       \textbf{SE} & E,LO,PE,PT,Z &            M \\
 \cite{Long19} &                           Pneumonia detection is improved by use of lung segmentation &       \textbf{SE} &           PM &   J,MO,PP,PR \\
 \cite{Hurt20} &                                                U-Net based model to segment pneumonia &       \textbf{SE} &           PM &           RP \\
\cite{Wang20f} &                        Multi-scale DenseNet based model for pneumothorax segmentation &       \textbf{SE} &           PT &           PR \\
 \cite{Liu20a} &     DenseNet based U-Net for segmentation of the left and right humerus of the infant &       \textbf{SE} &            Z &           PR \\
 \cite{Owai20} &        Uses a database of the intermediate ResNet-50 features to find similar studies &    \textbf{OT},IR &           TB &         MO,S \\
 \cite{Ouya20} &      Uses activation and gradient based attention for localization and classification &       \textbf{LC} &          C,X &            C \\
\cite{Raja20b} &                  Detects and localizes COVID-19 using various networks and ensembling &       \textbf{LC} &           CV & C,CC,PP,RP,X \\
 \cite{Sama21} &           GoogleNet trained with CXR patches, correlates with COVID-19 severity score &       \textbf{LC} &        CV,PM &         C,PR \\
 \cite{Park20} &     Proposes a segmentation and classification model compares with radiologist cohort &       \textbf{LC} &  LO,ND,PE,PT &           PR \\
  \cite{Nam19} & Trains a semisupervised network on a large CXR dataset with CT-confirmed nodule cases &       \textbf{LC} &           ND &           PR \\
 \cite{Pesc19} &    Defines a loss that minimizes the saliency map errors to improve model performance &       \textbf{LC} &           ND &           PR \\
 \cite{Tagh19} &     A weakly supervised localization with variational model, leverages attention maps &       \textbf{LC} &           PM &            C \\
  \cite{Li19b} &                      Attention guided CNN for pneumonia detection with bounding boxes &       \textbf{LC} &           PM &           RP \\
 \cite{Hwan19} &           A CNN for identification of abnormal CXRs and localization of abnormalities &       \textbf{LC} &            T &           PR \\
 \cite{Leni20} & Introduces a visualization method to identify regions of interest from classification &       \textbf{LC} &           TB &           PR \\
\cite{Hwan16a} &      Weakly supervised framework jointly trained with localization and classification &       \textbf{LC} &           TB &           PR \\
 \cite{Wang18} &                      Combines classification loss and autoencoder reconstruction loss &    \textbf{IG},SE &            T &     J,MO,O,S \\
 \cite{Seah19} &                     Wasserstein GAN to permute diseased radiographs to appear healthy &    \textbf{IG},LC &            Z &           PR \\
 \cite{Woll20} &       Novel GAN model trained with healthy and abnormal CXR to predict difference map &       \textbf{IG} &           PE &         SM,X \\
 \cite{Tang19} &  GANs with U-Net autoencoder and CNN discriminator and encoder for one-class learning &       \textbf{IG} &            T &            C \\
  \cite{Mao20} &   Autoencoder uses uncertainty for reconstruction error in one-class learning setting &       \textbf{IG} &            T &        PP,RP \\
 \cite{Leng20} &                          Continual learning methods to classify data from new domains &       \textbf{DA} &          C,M &          C,M \\
\cite{Tang19a} &           CycleGAN model to adapt adult to pediatric CXR for pneumonia classification &       \textbf{DA} &           PM &        PP,RP \\
 \cite{Gyaw19} &               Trains a Variational Autoencoder, uses encoded features to train models &                WS &            X &            X \\
 \cite{Gyaw20} &  Predicts labels for unlabeled data using latent space similarity for semisupervision &                WS &            X &            X \\
 \cite{McMa20} &                                   Y-Net to normalize image geometry for preprocessing &             SE,PR &           OR &        C,M,X \\
 \cite{Blai20} &                          COVID-19 opacity localization and severity detection on CXRs &             SE,LC &           CV &           PR \\
   \cite{Oh20} &         ResNet-18 backbone for Covid-19 classification with limited data availability &             SE,LC &        CV,PM &      CC,J,MO \\
 \cite{Tabi20} &     Proposes a new dataset COVIDGR and a novel method using transformations with GANs &             SE,IG &           CV &           CG \\
 \cite{Juni21} &                            DenseNet for cardiomegaly detection given lung cropped CXR &                SE &           CM &          O,P \\
 \cite{Tart20} &          Multiple models and combinations of CXR datasets used for COVID-19 detection &                SE &           CV &   C,CC,PR,RP \\
 \cite{Kusa21} &    ResNet-101 trained for COVID-19, heatmaps are generated for lung-segmented regions &                SE &        CV,PM &           PR \\
 \cite{Nara20} & Multiple architectures considered for two-stage classification of pediatric pneumonia &                SE &           PM &           PP \\
\cite{Raja19a} &                             Compares visualization methods for pneumonia localization &                SE &           PM &           PP \\
 \cite{Blum18} &              Classifies patches and uses the positive area size to classify the image &                SE &           PT &           PR \\
\cite{Raja18a} &                             Feature extraction from CNN models and ensembling methods &                SE &           TB &      MO,PR,S \\
 \cite{Subr19} &               Detection of central venous catheters using segmentation shape analysis &                SE &           TU &            C \\
 \cite{Mans16} &                Detection of air-trapping in pediatric CXRs using Stacked Autoencoders &                SE &            Z &           PR \\
\cite{Wang20a} &   Pneumoconiosis detection using Inception-v3 and evaluation against two radiologists &                SE &            Z &           PR \\
 \cite{Irvi19} &     Introduces CheXpert dataset and model performance on radiologist labeled test set &             RP,LC &            X &            X \\
   \cite{Oh19} &  Curates data for interval change detection, proposes method comparing local features &             RP,IC &            Z &           PR \\
 \cite{Dani19} &                  Parses reports to define a topic model and predicts those using CXRs &                RP &      CM,PE,Z &            O \\
 \cite{Chau20} &                Trains model using image and reports to improve image only performance &                RP &            E &            M \\
 \cite{Kara19} &  Extracts ambiguity of labels from reports, proposes model that uses this information &                RP &       E,PT,Z &            M \\
 \cite{Syed19} &  Creates and parses reports for ChestX-ray14 AP data to obtain 73 labels for training &                RP &           MN &            C \\
 \cite{Lase18} &                   Obtains findings by tagging common report sentences to train models &                RP &           MN &           PR \\
 \cite{Anna19} &            An ensemble of two CNNs to predict priority level for CXR queue management &                RP &            T &           PR \\
\cite{Balt19a} &        Evaluates bone suppression and lung segmentation, detection of 8 abnormalities &             PR,SE &   CM,PE,PT,Z &            O \\
 \cite{Ferr20} &          Classification of pediatric pneumonia types using adaped VGG-16 architecture &             PR,SE &           PM &           PP \\
 \cite{Vidy19} &   Evaluates various image preprocessing algorithms on the performance of DenseNet-121 &                PR &            T &    C,MO,PR,S \\
 \cite{Balt20} &         Detects 8 findings and analyzes how these can improve workflow prioritization &                OT &   CM,PE,PT,Z &          C,O \\
 \cite{Herm20} &                Proposes a model for weakly supervised classification and localization &             LC,WS &            C &            C \\
 \cite{Saed20} &                 Proposes a recurrent attention mechanism to improve model performance &                LC &            C &            C \\
 \cite{Balt19} &                  Evaluates the use of various model configurations for classification &                LC &            C &            C \\
  \cite{Cai18} &      Attention mining and knowledge preservation for classification with localization &                LC &            C &            C \\
 \cite{Wang20} &                          Attention based model compared with well-known architectures &                LC &            C &            C \\
   \cite{Ma19} &                      Minimizes the encoding differences of a CXR from multiple models &                LC &            C &          C,X \\
 \cite{Cohe20} &                  DenseNet used to predict COVID-19 severity as scored by radiologists &                LC &           CV &           CC \\
 \cite{Wang21} &     Uses a ResNet-50 backed segmentation model to detect healthy, pneumonia, COVID-19 &                LC &        CV,PM &        CC,RP \\
 \cite{Schw20} &                     Uses multi instance learning for classification with localization &                LC &      E,PM,PT &      M,PR,RP \\
  \cite{Yoo20} &                                     Lung cancer and nodule prediction using ResNet-34 &                LC &        LC,ND &           PR \\
 \cite{Kash20} &      GradCam based attention mining loss, compared with labels extracted from reports &                LC &           LO &            M \\
 \cite{Majk20} &            Trains Xception using >750k CXRs, compares results with radiologist labels &                LC &   LO,ND,PT,Z &         C,PR \\
\midrule
\multicolumn{5}{r}{\footnotesize\textit{continued on the next page}}\\
\bottomrule
\end{tabular}
\end{table*}

\begin{table*}[h!]
\ContinuedFloat
\centering
\scriptsize
\caption{Image-Level Prediction Studies (Section \ref{sec:studies_ilp}).\\ 
\textbf{Tasks}: AA=Adversarial Attack, DA=Domain Adaptation, IC=Interval Change, IG=Image Generation, IR=Image Retrieval, LC=Localization, OT=Other, PR=Preprocessing, RP=Report Parsing, SE=Segmentation, WS=Weak Supervision. \textbf{Bold font} in tasks implies that this additional task is central to the work and the study also appears in another table in this paper.\\ 
\textbf{Labels}: C=ChestX-Ray14, CM=Cardiomegaly, CV=COVID, E=Edema, GA=Gender/Age, L=Lung, LC=Lung Cancer, LO=Lesion or Opacity, M=MIMIC-CXR, MN=Many, ND=Nodule, OR=Orientation, P=PadChest, PE=Effusion, PL=PLCO, PM=Pneumonia, PT=Pneumothorax, Q=Image Quality, T=Triage/Abnormal, TB=Tuberculosis, TU=Catheter or Tube, X=CheXpert, Z=Other. \\ 
\textbf{Datasets}: BL=Belarus, C=ChestX-ray14, CC=COVID-CXR, CG=COVIDGR, J=JSRT+SCR, M=MIMIC-CXR, MO=Montgomery, O=Open-i, P=PadChest, PL=PLCO, PP=Ped-pneumonia, PR=Private, RP=RSNA-Pneumonia, S=Shenzen, SI=SIIM-ACR, SM=Simulated CXR from CT, X=CheXpert.}
\begin{tabular}{p{3.6cm}p{8.75cm}p{1.2cm}p{1.35cm}p{1.35cm}}
\toprule
\multicolumn{5}{r}{\footnotesize\textit{continued from the previous page}}\\
\midrule
      Citation &                                                                                Method & Other Tasks &       Labels &    Datasets \\
\midrule
 \cite{Hosc20} &                                  ResNet and VGG used to distinguish AP from PA images &          LC &           OR &       PR,RP \\
 \cite{Ruec20} &                   DenseNet-121 trained on public data evaluated using CT-based labels &          LC &        PE,PM &        C,PL \\
 \cite{Uret20} &   Evaluates the performance of various models trained on pediatric CXRs on adult CXRs &          LC &           PM &          PP \\
 \cite{Chen20} &                      Evaluates ensembling methods and visualization on pediatric CXRs &          LC &      PM,PT,Z &          PR \\
   \cite{Yi20} &     Compares a ResNet-152 against radiologists and shows the statistical significance &          LC &           PT &           C \\
\cite{Cros20a} &              Compares GradCAM with radiologist segmentations for evaluation of VGG-19 &          LC &           PT &        C,PR \\
\cite{Cros20b} &                 Apical regions and patches from them extracted to detect pneumothorax &          LC &           PT &          PR \\
 \cite{Tang20} &         Detection of abnormality, various networks compared with radiologist labeling &          LC &            T &   C,O,PP,RP \\
\cite{Raja20a} &            Evaluates pre-training on ImageNet and CheXpert on various models/settings &          LC &            T &          RP \\
 \cite{Naka21} &     Proposes a GAN-based model trained only with healthy images for anomaly detection &          LC &            T &          RP \\
 \cite{Pasa19} &                                  Proposes a new model for faster classification of TB &          LC &           TB &     BL,MO,S \\
\cite{Hwan19a} &             Evalutes the use of a ResNet based model on a large gold standard dataset &          LC &           TB &          PR \\
 \cite{Sing19} &                Evaluates multiple models for detection of feeding tube malpositioning &          LC &           TU &          PR \\
 \cite{Chak20} &                  Graph CNN solution with ensembling which models disease dependencies &          LC &            X &           X \\
 \cite{Mats20} &                   Curates a dataset of heart failure cases and evaluates VGG-16 on it &          LC &            Z &           C \\
   \cite{Su21} &                     CNN for identifiying the presence of subphrenic free air from CXR &          LC &            Z &          PR \\
  \cite{Zou20} &         Evaluates several models to predict hypertension and artery systolic pressure &          LC &            Z &          PR \\
 \cite{Zuck20} &  Uses ResNet-18 to measure the Brassfield Score, predicts Cystic Fibrosis based on it &          LC &            Z &          PR \\
 \cite{Camp18} &                            Simulates CXRs from CT scans and predicts emphysema scores &          LC &            Z &          PR \\
 \cite{Toba20} &      Inception network to predict pulmonary to systemic flow ratio from pediatric CXR &          LC &            Z &          PR \\
  \cite{Li20b} &                         Predicts COVID-19 severity by comparing CXRs to previous ones &       IC,LC &           CV &          PR \\
  \cite{Luo20} &                    Addresses domain and label discrepancies in multi-dataset training &          DA &       C,TB,X &      C,PR,X \\
  \cite{Xue19} &               Method to increase robustness of CNN classifiers to adversarial samples &       AA,IG &           PM &          RP \\
  \cite{Li20a} &      Uses the features extracted from the training dataset to detect adversarial CXRs &          AA &            C &           C \\
 \cite{Anan20} &               Self-supervision and adversarial training improves on transfer learning &          AA &           PM &          PP \\
 \cite{Khat21} &    Claims 0.99 AUC for predicting TB, uses complex feature engineering and ensembling &             &              &             \\
 \cite{Schr21} & ResNet model trained with frontal and lateral images to predict COPD with PFT results &             &              &          PR \\
\cite{Zhan21a} & One-class identification of viral pneumonia cases compared with binary classification &             &              &          PR \\
 \cite{Bala20} & A distributed learning method that overcomes problems of multi-institutional settings &             &            C &           C \\
 \cite{Burw19} &  Geometric deep learning including metadata with graph structure.  Application to CXR &             &            C &           C \\
 \cite{Nugr21} &                  Proposes a new weighting scheme to impove abnormality classification &             &            C &           C \\
 \cite{DSou19} &          ResNet-34 used with various training settings for multi-label classification &             &            C &           C \\
 \cite{Sira19} &  Investigates effect of data augmentations on classification with Inception-Resnet-v2 &             &            C &           C \\
  \cite{Mao18} &      Proposes a variational/generative architecture, demonstrates performance on CXRs &             &            C &           C \\
 \cite{Rajp18} &                    Evaluates the performance of an ensemble against many radiologists &             &            C &           C \\
 \cite{Kurm19} &                       Novel method for multi-label classification, application to CXR &             &            C &           C \\
 \cite{Paul20} &           Defines a few-shot learning method by extracting features from autoencoders &             &            C &           C \\
 \cite{Unni20} &                Mean teacher inspired a probablistic graphical model with a novel loss &             &            C &           C \\
 \cite{Mich20} &       Examines the effect of denoising on pathology classification using DenseNet-121 &             &            C &           C \\
\cite{Wang21b} &         Proposes integrating three attention mechanisms that work at different levels &             &            C &           C \\
 \cite{Paul21} &           Step-wise trained CNN and saliency-based autoeencoder for few shot learning &             &            C &         C,O \\
\cite{Paul21a} &   Uses CT and CXR reports with CXR images during training to diagnose unseen diseases &             &            C &        C,PR \\
 \cite{Bust20} &         Proposes a new dataset PadChest with multi-label labels and radiology reports &             &            C &           P \\
   \cite{Li21} &                   Lesion detection network used to improve image-level classification &             &            C &          PR \\
 \cite{Ghes19} &         Method to produce confidence measure alongside probability, uses DenseNet-121 &             &         C,PL &        C,PL \\
 \cite{Hagh20} &     Uses self-supervised learning for pretraining, compares with ImageNet pretraining &             &         C,PT &        C,SI \\
 \cite{Zhou20} &        Proposes a new CXR pre-training method, compares with pre-training on ImageNet &             &          C,X &      C,RP,X \\
\cite{Chen20b} &    Proposes a graph convolutional network framework which models disease dependencies &             &          C,X &         C,X \\
\cite{Zhou19a} &                             Compares several models for the detection of cardiomegaly &             &           CM &           C \\
 \cite{Boug20} &                      Tests four off-the-shelf networks for prediction of cardiomegaly &             &           CM &          PR \\
 \cite{Bres18} &     Inception v3 trained to detect 4 abnormalities and compared with expert observers &             &    CM,E,LO,Z &          PR \\
 \cite{Cice17} &       GoogLeNet to classify normal and 5 abnormalities on a large proprietary dataset &             & CM,E,PE,PT,Z &          PR \\
 \cite{Bar15a} & Compares the performance of deep learning with traditional feature extraction methods &             &        CM,PE &          PR \\
  \cite{Bar15} &          ImageNet pre-training and feature extraction methods for pathology detection &             &      CM,PE,Z &          PR \\
 \cite{Grin21} &                 An ensemble of DenseNet-121 networks used for COVID-19 classification &             &           CV &        C,PR \\
   \cite{Hu21} &      Investigates the value of soft tissue CXR for training DenseNet-121 for COVID-19 &             &           CV &     C,PR,RP \\
  \cite{Zhu20} &                                 Labels and predicts COVID-19 severity stage using CNN &             &           CV &          CC \\
 \cite{Fric21} & Uses a model trained on COVID-19 cases to evaluate the effect of an imaging parameter &             &           CV &          PR \\
 \cite{Wehb20} & COVID-19 detection based on RT-PCR labels, evaluates an ensamble against radiologists &             &           CV &          PR \\
 \cite{Cast21} &                                      Ensemble of ResNet models for COVID-19 detection &             &           CV &          PR \\
\cite{Zhan21c} &                   Compares the performance of a DenseNet-121 ensemble to radiologists &             &        CV,PM &          PR \\
\cite{Wang19b} &        Various models and use of semi-supervised labels for edema severity estimation &             &            E &           M \\
\cite{Kara19a} &                                  Age prediction on PA or AP images using DenseNet-169 &             &           GA &           C \\
 \cite{Xue18c} & Gender prediction using features from deep-learning models in traditional classifiers &             &           GA & J,MO,O,PR,S \\
 \cite{Sabo20} &                         Age prediction on  AP images using DenseNet-121 and ResNet-50 &             &           GA &           X \\
  \cite{Lu20a} &         Combines the CXR with age/sex/smoking history to predict the lung cancer risk &             &           LC &          PL \\
 \cite{Tham20} &                  Densenet-121 pre-trained with public data used to identify 6 classes &             &    LC,T,TB,Z &          PR \\
  \cite{Kuo21} &               Evaluates deep learning on pictures of CXRs captured with mobile phones &             &          M,X &         M,X \\
\midrule
\multicolumn{5}{r}{\footnotesize\textit{continued on the next page}}\\
\bottomrule
\end{tabular}
\end{table*}

\begin{table*}[h!]
\ContinuedFloat
\centering
\scriptsize
\caption{Image-Level Prediction Studies (Section \ref{sec:studies_ilp}).\\ 
\textbf{Tasks}: AA=Adversarial Attack, DA=Domain Adaptation, IC=Interval Change, IG=Image Generation, IR=Image Retrieval, LC=Localization, OT=Other, PR=Preprocessing, RP=Report Parsing, SE=Segmentation, WS=Weak Supervision. \textbf{Bold font} in tasks implies that this additional task is central to the work and the study also appears in another table in this paper.\\ 
\textbf{Labels}: C=ChestX-Ray14, CM=Cardiomegaly, CV=COVID, E=Edema, GA=Gender/Age, L=Lung, LC=Lung Cancer, LO=Lesion or Opacity, M=MIMIC-CXR, MN=Many, ND=Nodule, OR=Orientation, P=PadChest, PE=Effusion, PL=PLCO, PM=Pneumonia, PT=Pneumothorax, Q=Image Quality, T=Triage/Abnormal, TB=Tuberculosis, TU=Catheter or Tube, X=CheXpert, Z=Other. \\ 
\textbf{Datasets}: BL=Belarus, C=ChestX-ray14, CC=COVID-CXR, CG=COVIDGR, J=JSRT+SCR, M=MIMIC-CXR, MO=Montgomery, O=Open-i, P=PadChest, PL=PLCO, PP=Ped-pneumonia, PR=Private, RP=RSNA-Pneumonia, S=Shenzen, SI=SIIM-ACR, SM=Simulated CXR from CT, X=CheXpert.}
\begin{tabular}{p{3.6cm}p{8.75cm}p{1.2cm}p{1.35cm}p{1.35cm}}
\toprule
\multicolumn{5}{r}{\footnotesize\textit{continued from the previous page}}\\
\midrule
      Citation &                                                                                Method & Other Tasks & Labels &     Datasets \\
\midrule
\cite{Cohe20c} &        Investigates the domain and label shift across publicly available CXR datasets &             &     MN & C,M,O,P,RP,X \\
 \cite{Hash20} &    Explores the use of the lateral view CXR for classification of 64 different labels &             &   MN,P &            P \\
 \cite{Rajk17} &             Classification of CXRs as Frontal or Lateral using GoogLeNet architecture &             &     OR &           PR \\
 \cite{Cros19} &                       Assesses the effect of imprinted labels on AP/PA classification &             &     OR &           PR \\
 \cite{Cros20} &    Distinguishes the CXR orientation, bone CXRs and soft tissue CXRs from dual energy &             &   OR,Z &           PR \\
\cite{Bert19a} &                  Compares PA and Lateral images for pathology detection with DenseNet &             &      P &            P \\
\cite{Chen19a} &     Introduces a loss term that uses the label hierarchy to improve model performance &             &     PL &           PL \\
 \cite{Shah20} &                                                Trains VGG-16 on Ped-pneumonia dataset &             &     PM &           PP \\
   \cite{Qu20} &            Methods to mitigate imbalanced class sizes. Applied to CXR using ResNet-18 &             &     PM &           PP \\
  \cite{Yue20} &                         Evaluation of MobileNet to detect pneumonia on pediatric CXRs &             &     PM &           PP \\
 \cite{Elsh20} &                               Compares multiple architectures for pneumonia detection &             &     PM &           PP \\
 \cite{Mitt20} &     Evaluates various capsule network architectures for pediatric pneumonia detection &             &     PM &           PP \\
 \cite{Long20} &                                       Uses ResNet-50 to classify paediatric pneumonia &             &     PM &           PR \\
 \cite{Gane19} &              Compares traditional and generative data augmentation techniques on CXRs &             &     PM &           RP \\
 \cite{Ravi19} & Addresses catastrophic forgetting, application to pneumothorax detection using VGG-13 &             &     PT &            C \\
 \cite{Tayl18} &   Construction of large dataset, multiple architectures and hyperparameters optimized &             &     PT &           PR \\
 \cite{Kita20} &          Model pre-trained with public data and fine-tuned for pneumothorax detection &             &     PT &           PR \\
 \cite{Mora19} &                       DenseNet-121 used to detect CXRs with acquisition-based defects &             &      Q &            C \\
 \cite{Taka20} &            GoogleNet combined with rule-based approach to determine the image quality &             &      Q &           PR \\
  \cite{Pan19} &    Detects abnormal CXRs using several models.  Evaluates on independent private data &             &      T &            C \\
\cite{Mora19a} &    Defines a model on top of features extracted from Inception-ResNet-v2 for triaging &             &      T &            C \\
 \cite{Wong20} &           Collects features from pretrained models and adds a CNN on top for triaging &             &      T &          C,M \\
 \cite{Jang20} &  Studies the effect of various label noise levels on classification with DenseNet-121 &             &      T &       C,PR,X \\
 \cite{Dunn19} & Various models for detection of abnormal CXRs, effect of different training set sizes &             &      T &           PR \\
  \cite{Nam20} &     Defines 10 abnormalities to define a triaging model and uses CT based test labels &             &      T &           PR \\
 \cite{Dyer21} &                 Ensembe of DenseNet and EffecientNet for identification of normal CXR &             &      T &           PR \\
 \cite{Ogaw19} &                           Examines the use of data augmentation in small data setting &             &      T &           PR \\
 \cite{Elli20} &           Evaluation of extra supervision in the form of localized region of interest &             &      T &           PR \\
 \cite{Raja19} &                  Evalutates various models and ensembling methods for the triage task &             &      T &           RP \\
\cite{Lakh17a} &                         Evaluates deep learning approaches for tuberculosis detection &             &     TB &   BL,MO,PR,S \\
 \cite{Hwan16} &                     Evaluates the use of transfer learning for tuberculosis detection &             &     TB &      MO,PR,S \\
 \cite{Siva18} &            Extracts feaures using off-the-shelf models and trains a model using those &             &     TB &      MO,PR,S \\
 \cite{Ayaz21} &                     Combines hand-crafted features and CNN for tuberculosis diagnosis &             &     TB &         MO,S \\
  \cite{UlA20} &                                    Evaluates a Bayesian-based CNN for detection of TB &             &     TB &         MO,S \\
 \cite{Rajp20} &     Evaluates assisting clinicians with an AI based system to improve diagnosis of TB &             &     TB &           PR \\
  \cite{Heo19} &          Various architectures, inclusion of patient demographics in model considered &             &     TB &           PR \\
  \cite{Kim18} &   Addresses preservation of learned data, application to TB detection using ResNet-21 &             &     TB &           PR \\
 \cite{Goze19} &     Pre-training using CXR pathology and metadata labels, application to TB detection &             &     TB &            S \\
 \cite{Raja20} &           Compares various models using various pretraining and ensembling strategies &             &     TB &            S \\
 \cite{Lakh17} &      Evaluates models on detecting the position of feeding tube in abdominal and CXRs &             &     TU &           PR \\
 \cite{Mitr20} &    Comparison of seven architectures and ensembling for detection of nine pathologies &             &      X &            X \\
 \cite{Pham20} & A method to incorporate label dependencies and uncertainty data during classification &             &      X &            X \\
 \cite{Raja21} &                Proposes self-training and student-teacher model for sample effeciency &             &      X &            X \\
 \cite{Call19} &                      Analyses the effect of label noise in training and test datasets &             &      Z &            C \\
 \cite{Desh20} &       Labels 6 different foreign object types and detects using various architectures &             &      Z &            M \\
   \cite{Lu19} &           Evaluates the use of CXRs to predict long term mortality using Inception-v4 &             &      Z &           PL \\
 \cite{Zhan21} &   Low-res segmentation is used to crop high-res lung areas and predict pneumoconiosis &             &      Z &           PR \\
 \cite{Devn21} &    Pneumoconiosis prediction with DenseNet-121 and SVMs applied to extracted features &             &      Z &           PR \\
  \cite{Liu17} &            Detection of coronary artery calcification using various CNN architectures &             &      Z &           PR \\
 \cite{Hira21} & ResNet-50 for detection of the presence of elevated pulmonary arterial wedge pressure &             &      Z &           PR \\
 \cite{Kusu20} &    A network is designed to identify subjects with elevated pulmonary artery pressure &             &      Z &        PR,RP \\
\bottomrule
\end{tabular}
\end{table*}

%% file: tables/Segmentation.tex
\begin{table*}[h!]
\centering
\scriptsize
\caption{Segmentation Studies (Section \ref{sec:studies_seg}).\\ 
\textbf{Tasks}: DA=Domain Adaptation, IG=Image Generation, IL=Image-level Predictions, LC=Localization, PR=Preprocessing, WS=Weak Supervision. \textbf{Bold font} in tasks implies that this additional task is central to the work and the study also appears in another table in this paper.\\ 
\textbf{Labels}: C=ChestX-Ray14, CL=Clavicle, CM=Cardiomegaly, CV=COVID, E=Edema, H=Heart, L=Lung, LO=Lesion or Opacity, PE=Effusion, PM=Pneumonia, PT=Pneumothorax, R=Rib, TU=Catheter or Tube, Z=Other. \\ 
\textbf{Datasets}: BL=Belarus, C=ChestX-ray14, J=JSRT+SCR, M=MIMIC-CXR, MO=Montgomery, O=Open-i, PP=Ped-pneumonia, PR=Private, RP=RSNA-Pneumonia, S=Shenzen, SI=SIIM-ACR, SM=Simulated CXR from CT.}
\label{tab:segmentation_table}
\begin{tabular}{p{3.6cm}p{8.75cm}p{1.2cm}p{1.35cm}p{1.35cm}}
\toprule
      Citation &                                                                                Method &       Other Tasks &       Labels &   Datasets \\
\midrule
   \cite{Yu20} &           A model based on U-Net and Faster R-CNN to detect PICC catether and its tip &    \textbf{LC},PR &           TU &         PR \\
\cite{Zhan19b} &                       Tailored Mask R-CNN for simultaneous detection and segmentation &       \textbf{LC} &            L &         PR \\
 \cite{Wess19} &                               Uses Mask R-CNN iteratively to segment and detect ribs. &       \textbf{LC} &            R &         PR \\
  \cite{Liu19} &          Combines lung cropped CXR model and a CXR model to improve model performance & \textbf{IL},LC,PR &          C,L &        C,J \\
 \cite{Soga20} &         Comparison of image-level prediction and segmentation models for cardiomegaly &       \textbf{IL} &           CM &          C \\
 \cite{Que18a} &                  A network with DenseNet and U-Net for classification of cardiomegaly &       \textbf{IL} &           CM &          C \\
   \cite{Li19} &            U-Net based model for heart and lung segmentation for cardiothoracic ratio &       \textbf{IL} &           CM &         PR \\
 \cite{Mora20} &        Combines lung cropped CXR model and a CXR model using the segmentation quality &       \textbf{IL} & E,LO,PE,PT,Z &          M \\
 \cite{Long19} &                           Pneumonia detection is improved by use of lung segmentation &       \textbf{IL} &           PM & J,MO,PP,PR \\
 \cite{Hurt20} &                                                U-Net based model to segment pneumonia &       \textbf{IL} &           PM &         RP \\
\cite{Wang20f} &                        Multi-scale DenseNet based model for pneumothorax segmentation &       \textbf{IL} &           PT &         PR \\
 \cite{Liu20a} &     DenseNet based U-Net for segmentation of the left and right humerus of the infant &       \textbf{IL} &            Z &         PR \\
\cite{Tang19b} &               Attention-based network and CXR synthesis process for data augmentation &    \textbf{IG},IG &            L &    J,MO,PR \\
\cite{Esla20a} &            Conditional GANs for multi-class segmentation of heart,clavicles and lungs &       \textbf{IG} &       CL,H,L &          J \\
 \cite{Onod20} &    Processing method to produce scatter-corrected CXRs and segments masses with U-Net &       \textbf{IG} &           LO &         SM \\
\cite{Oliv20b} &                                            MUNIT based DA model for lung segmentation &       \textbf{DA} &       CL,H,L &          J \\
 \cite{Dong18} &                            Adversarial training of lung and heart segmentation for DA &       \textbf{DA} &           CM &       J,PR \\
 \cite{Zhan18} &       CycleGAN guided by a segmentation module to convert CXR to CT projection images &       \textbf{DA} &        H,L,Z &         PR \\
\cite{Chen18a} &                CycleGAN based DA model with semantic aware loss for lung segmentation &       \textbf{DA} &            L &         MO \\
\cite{Oliv20a} &                                       Conditional GANs based DA for bone segmentation &       \textbf{DA} &            R &         SM \\
 \cite{Shah18} &     FCN based novel model incorporating weak landmarks and bounding boxes annotations &                WS &       CL,H,L &          J \\
 \cite{Bort19} &          U-Net segmentation model integrating unlabeled data through consistency loss &                WS &       CL,H,L &          J \\
 \cite{Ouya19} &     Attention masks derived from classification model to guide the segmentation model &                IL &           PT &         PR \\
 \cite{Frid19} &           U-Net based network for classification and segmentation with simulated data &                IL &            Z &        C,J \\
 \cite{Sull20} &        U-Net based model for segmentation and a classification for existance of lines &                IL &            Z &         PR \\
 \cite{Card21} &                U-Net for bone suppression given lung-segmented CXR image with patches &                   &              &         PR \\
\cite{Zhan20a} &                      Proposes teacher-student based learning with noisy segmentations &                   &       CL,H,L &          J \\
 \cite{Khol20} &     Various FCN based models explored for simultaneous pixel and contour segmentation &                   &       CL,H,L &          J \\
 \cite{Novi18} &     Investigates various FCN type architecture including U-Net for organ segmentation &                   &       CL,H,L &          J \\
 \cite{Bonh19} &                           Capsule networks adapted for multi-class organ segmentation &                   &       CL,H,L &          J \\
 \cite{Arsa20} &             U-Net based architecture with residual connections for organ segmentation &                   &       CL,H,L &     J,MO,S \\
\cite{Wang20b} &                                   U-Net based architecture based on dense connections &                   &         CL,R &         PR \\
 \cite{Mort21} &          CNN trained with CT projection images for quantification of airspace disease &                   &           CV &         PR \\
 \cite{Larr20} &                     Denoising autoencoder as post-processing to improve segmentations &                   &          H,L &          J \\
 \cite{Hols20} &        Evaluates U-Net performance with various loss functions, and data augmentation &                   &          H,L &         PR \\
\cite{Mans20a} &          Stacked denoising autoencoder model for space and shape parameter estimation &                   &            L &    BL,J,PR \\
 \cite{Amir20} &         Investigates the effect of fine-tuning different layers for U-Net based model &                   &            L &          J \\
   \cite{Lu20} &                               Proposes a human-in-the-loop one shot anatomy segmentor &                   &            L &          J \\
  \cite{Li21c} &             U-Net with conditional random field post processing for lung segmentation &                   &            L &          J \\
 \cite{Port20} &                               Investigates U-Net with different optimizer and dropout &                   &            L &          J \\
 \cite{Yahy20} &    U-Net with dense connections for reducing network parameters for lung segmentation &                   &            L &       J,MO \\
  \cite{Kim21} &                                       U-Net with self attention for lung segmentation &                   &            L &     J,MO,S \\
 \cite{Arba17} &                                      Multi-scale and patch-based CNN to segment lungs &                   &            L &       J,PR \\
 \cite{Rahm21} &                      U-Net based model for lung segmentation trained with CXR patches &                   &            L &         MO \\
 \cite{Souz19} &                         Two stage patch based CNN for refined lung field segmentation &                   &            L &         MO \\
 \cite{Mill18} &                Encoder-decoder architecture with ConvLSTM and ResNet for segmentation &                   &            L &         MO \\
\cite{Zhan19a} &       Encoder-decoder based CNN with novel edge guidance module for lung segmentation &                   &            L &         MO \\
 \cite{Math19} &  Proposes a convolutional LSTM model for ultrasound, uses CXR as a secondary modality &                   &            L &         MO \\
 \cite{Kita19} &                                       U-Net based segmentation model for dynamic CXRs &                   &            L &         PR \\
 \cite{Furu19} &               U-Net for whole lung region segmentation including where heart overlaps &                   &            L &         PR \\
  \cite{Xue20} &                     Cascaded U-net with sample selection with imperfect segmentations &                   &            L &          S \\
\cite{Wang20g} &            ResNet-50 based architecture with segmentation and classification branches &                   &           PT &         SI \\
 \cite{Tolk20} &       Investigates U-Net based models with various backbone encoders for pneumothorax &                   &           PT &         SI \\
 \cite{Groz20} &           Ensemble of three LinkNet based networks and with multi-step postprocessing &                   &           PT &         SI \\
  \cite{Xue18} &                       Cascaded network with Faster R-CNN and U-Net for aortic knuckle &                   &            Z &          J \\
  \cite{Yi20a} &               Multi-scale U-Net based model with recurrent module for foreign objects &                   &            Z &          O \\
  \cite{Lee18} &            Two FCN to segment peripherally inserted central catheter line and its tip &                   &            Z &         PR \\
 \cite{Pan19a} & Two Mask R-CNN to segment the spine and vertebral bodies and calculate the Cobb angle &                   &            Z &         PR \\
\bottomrule
\end{tabular}
\end{table*}

%% file: tables/Localization.tex
\begin{table*}[h!]
\centering
\scriptsize
\caption{Localization Studies (Section \ref{sec:studies_loc}).\\ 
\textbf{Tasks}: IC=Interval Change, IL=Image-level Predictions, PR=Preprocessing, RP=Report Parsing, SE=Segmentation, WS=Weak Supervision. \textbf{Bold font} in tasks implies that this additional task is central to the work and the study also appears in another table in this paper.\\ 
\textbf{Labels}: C=ChestX-Ray14, CM=Cardiomegaly, CV=COVID, L=Lung, LC=Lung Cancer, LO=Lesion or Opacity, ND=Nodule, PE=Effusion, PM=Pneumonia, PT=Pneumothorax, R=Rib, T=Triage/Abnormal, TB=Tuberculosis, TU=Catheter or Tube, X=CheXpert, Z=Other. \\ 
\textbf{Datasets}: C=ChestX-ray14, CC=COVID-CXR, J=JSRT+SCR, M=MIMIC-CXR, O=Open-i, PP=Ped-pneumonia, PR=Private, RP=RSNA-Pneumonia, S=Shenzen, X=CheXpert.}
\label{tab:localization_table}
\begin{tabular}{p{3.6cm}p{8.75cm}p{1.2cm}p{1.35cm}p{1.35cm}}
\toprule
      Citation &                                                                                Method &    Other Tasks &        Labels &     Datasets \\
\midrule
   \cite{Yu20} &           A model based on U-Net and Faster R-CNN to detect PICC catether and its tip & \textbf{SE},PR &            TU &           PR \\
\cite{Zhan19b} &                       Tailored Mask R-CNN for simultaneous detection and segmentation &    \textbf{SE} &             L &           PR \\
 \cite{Wess19} &                               Uses Mask R-CNN iteratively to segment and detect ribs. &    \textbf{SE} &             R &           PR \\
 \cite{Ouya20} &      Uses activation and gradient based attention for localization and classification &    \textbf{IL} &           C,X &            C \\
\cite{Raja20b} &                  Detects and localizes COVID-19 using various networks and ensembling &    \textbf{IL} &            CV & C,CC,PP,RP,X \\
 \cite{Sama21} &           GoogleNet trained with CXR patches, correlates with COVID-19 severity score &    \textbf{IL} &         CV,PM &         C,PR \\
 \cite{Park20} &     Proposes a segmentation and classification model compares with radiologist cohort &    \textbf{IL} &   LO,ND,PE,PT &           PR \\
  \cite{Nam19} & Trains a semisupervised network on a large CXR dataset with CT-confirmed nodule cases &    \textbf{IL} &            ND &           PR \\
 \cite{Pesc19} &    Defines a loss that minimizes the saliency map errors to improve model performance &    \textbf{IL} &            ND &           PR \\
 \cite{Tagh19} &     A weakly supervised localization with variational model, leverages attention maps &    \textbf{IL} &            PM &            C \\
  \cite{Li19b} &                      Attention guided CNN for pneumonia detection with bounding boxes &    \textbf{IL} &            PM &           RP \\
 \cite{Hwan19} &           A CNN for identification of abnormal CXRs and localization of abnormalities &    \textbf{IL} &             T &           PR \\
 \cite{Leni20} & Introduces a visualization method to identify regions of interest from classification &    \textbf{IL} &            TB &           PR \\
\cite{Hwan16a} &      Weakly supervised framework jointly trained with localization and classification &    \textbf{IL} &            TB &           PR \\
\cite{Chen20a} &              Extract nodule candidates using traditional methods and trains GoogleNet &          SE,PR &            ND &            J \\
 \cite{Schu20} &                       RetinaNet for detecting nodules incorporating lung segmentation &             SE &            ND &         J,PR \\
  \cite{Tam20} &       Combines reports and CXRs for weakly supervised localization and classification &          RP,WS &         PM,PT &          C,M \\
\cite{Mora18a} &           Proposes a model using LSTM and CNN, combining reports and images as inputs &          IL,RP &         CM,ND &          C,O \\
 \cite{Khak19} &   Adversarially trained weakly supervised localization framework for interpretability &             IL &             C &            C \\
  \cite{Kim20} & Evaluates the effect of image size for nodule detection with Mask R-CNN and RetinaNet &             IL &            ND &           PR \\
  \cite{Cho20} &     Evaluates the reproducibility of YOLO for disease localization in follow up exams &             IC & LO,ND,PE,PT,Z &           PR \\
  \cite{Kim19} &  Evaluates the reproducability of various detection architectures in follow up exams. &             IC &            ND &           PR \\
 \cite{Cha19a} &        ResNet model using CT and surgery based annotations for lung cancer prediction &                &            LC &           PR \\
 \cite{Take19} &                                                R-CNN for localization of lung nodules &                &            ND &            J \\
 \cite{Wang17} &          Fuses AlexNet and hand-crafted features to improve random forest performance &                &            ND &            J \\
   \cite{Li20} &             Patch-based nodule detectin, combines features from different resolutions &                &            ND &         J,PR \\
 \cite{Park19} &                Evaluates the detection of pneumothorax before, 3h and 1d after biopsy &                &            PT &           PR \\
 \cite{Made18} &              Proposes a U-Net based model for localizing and labeling individual ribs &                &             R &            O \\
 \cite{Xue18a} &                         AlexNet for localizing tuberculosis with patch-based approach &                &            TB &            S \\
 \cite{Berg20} &                                 Localizes anatomical features for image quality check &                &             Z &           PR \\
\bottomrule
\end{tabular}
\end{table*}

%% file: tables/Image-Generation.tex
\begin{table*}[h!]
\centering
\scriptsize
\caption{Image Generation Studies (Section \ref{sec:studies_gen}).\\ 
\textbf{Tasks}: DA=Domain Adaptation, IC=Interval Change, IG=Image Generation, IL=Image-level Predictions, LC=Localization, PR=Preprocessing, RE=Registration, SE=Segmentation, SR=Super Resolution. \textbf{Bold font} in tasks implies that this additional task is central to the work and the study also appears in another table in this paper.\\ 
\textbf{Labels}: BS=Bone Suppression, C=ChestX-Ray14, CL=Clavicle, CM=Cardiomegaly, CV=COVID, E=Edema, H=Heart, L=Lung, LO=Lesion or Opacity, PE=Effusion, PT=Pneumothorax, T=Triage/Abnormal, TB=Tuberculosis, Z=Other. \\ 
\textbf{Datasets}: C=ChestX-ray14, CC=COVID-CXR, J=JSRT+SCR, MO=Montgomery, O=Open-i, PL=PLCO, PP=Ped-pneumonia, PR=Private, RP=RSNA-Pneumonia, S=Shenzen, SM=Simulated CXR from CT, X=CheXpert.}
\label{tab:image_generation_table}
\begin{tabular}{p{3.6cm}p{8.75cm}p{1.2cm}p{1.35cm}p{1.35cm}}
\toprule
      Citation &                                                                                Method &       Other Tasks &     Labels & Datasets \\
\midrule
\cite{Tang19b} &               Attention-based network and CXR synthesis process for data augmentation &    \textbf{SE},IG &          L &  J,MO,PR \\
\cite{Esla20a} &            Conditional GANs for multi-class segmentation of heart,clavicles and lungs &       \textbf{SE} &     CL,H,L &        J \\
 \cite{Onod20} &    Processing method to produce scatter-corrected CXRs and segments masses with U-Net &       \textbf{SE} &         LO &       SM \\
 \cite{Wang18} &                      Combines classification loss and autoencoder reconstruction loss &    \textbf{IL},SE &          T & J,MO,O,S \\
 \cite{Seah19} &                     Wasserstein GAN to permute diseased radiographs to appear healthy &    \textbf{IL},LC &          Z &       PR \\
 \cite{Woll20} &       Novel GAN model trained with healthy and abnormal CXR to predict difference map &       \textbf{IL} &         PE &     SM,X \\
 \cite{Tang19} &  GANs with U-Net autoencoder and CNN discriminator and encoder for one-class learning &       \textbf{IL} &          T &        C \\
  \cite{Mao20} &   Autoencoder uses uncertainty for reconstruction error in one-class learning setting &       \textbf{IL} &          T &    PP,RP \\
 \cite{Maha19} &           Conditional GAN based DA for image registration using segmentation guidance & \textbf{DA},RE,SE &          L &        C \\
 \cite{Mada18} &          Adversarial based method adapting new domains for abnormality classification &    \textbf{DA},IL &         CM &       PL \\
 \cite{Umeh17} &                                    Proposes a patch-based CNN super resolution method &                SR &          Z &        J \\
 \cite{Uzun19} &                    Generates high resolution CXRs using multi-scale, patch based GANs &                SR &          Z &        O \\
 \cite{Zhan19} &        Novel GAN model with sketch guidance module for high resolution CXR generation &                SR &          Z &       PP \\
  \cite{Lin20} &  AutoEncoder for bone suppression and segmentation with statistical similarity losses &             SE,PR &         BS &        J \\
 \cite{Dong19} &              Uses neural architecture search to find a discriminator network for GANs &                SE &        H,L &     J,PR \\
\cite{Tagh19a} &     Proposes an iterative gradient based input preprocessing for improved performance &                SE &          L &        S \\
 \cite{Fang20} &  Learns transformations to register two CXRs, uses the difference for interval change &             RE,IC &          Z &       PR \\
 \cite{Yang17} &                         Generates bone and soft tissue (dual energy) images from CXRs &                PR &         BS &       PR \\
 \cite{Zars19} &       Proposes an CNN with multi-resolution decomposition for bone suppression images &                PR &         BS &       PR \\
 \cite{Goze20} &         U-Net for bone generation with CT projection images, used for CXR enhancement &                PR &         BS &       SM \\
  \cite{Lee19} &                                       U-Net based network to generate dual energy CXR &                PR &          Z &       PR \\
  \cite{Liu20} &        GAN integrates edges of ribs and clavicles to guide DES-like images generation &                PR &          Z &       PR \\
 \cite{Xing19} & Generates diseased CXRs, evaluates their realness with radiologists and trains models &                LC &          C &        C \\
  \cite{Li19a} &       Novel CycleGAN model to decompose CXR images incorporating CT projection images &                IL &          C &  C,PR,SM \\
 \cite{Sale19} &             Uses DCGAN model to generate CXR with abnormalities for data augmentation &                IL & CM,E,PE,PT &       PR \\
 \cite{Alba17} &     U-Net based architecture to decompose CXR structures, application to TB detection &                IL &         TB &       PR \\
 \cite{Mora18} &               Two DCGAN trained with normal and abnormal images for data augmentation &                IL &          Z &       PL \\
 \cite{Bigo19} &  Novel conditional GAN using lung function test results to visualize COPD progression &                IL &          Z &       PR \\
 \cite{Zare21} &          Conditional GAN and two variational autoencoders designed for CXR generation &                   &            &       PR \\
 \cite{Gomi20} &                                    Novel reconstruction algorithm for CXR enhancement &                   &            &       PR \\
\cite{Zhou20a} &             Bone shadow suppression using conditional GANS with dilated U-Net variant &                   &         BS &        J \\
\cite{Mats20a} &                              Generates CXRs from CT to train CNN for bone suppression &                   &         BS &       PR \\
 \cite{Zuna21} &             Generates COVID-19 CXR images to improve network training and performance &                   &         CV &    CC,RP \\
 \cite{Baya20} &      2D-to-3D encoder-decoder network for generating 3D spine models from CXR studies &                   &          Z &       PR \\
 \cite{Bigo20} &        Generates normal from abnormal CXRs, uses the deformations as disease evidence &                   &          Z &       PR \\
\bottomrule
\end{tabular}
\end{table*}

%% file: tables/Domain-Adaptation.tex
\begin{table*}[h!]
\centering
\scriptsize
\caption{Domain Adaptation Studies (Section \ref{sec:studies_da}).\\ 
\textbf{Tasks}: IG=Image Generation, IL=Image-level Predictions, RE=Registration, SE=Segmentation. \textbf{Bold font} in tasks implies that this additional task is central to the work and the study also appears in another table in this paper.\\ 
\textbf{Labels}: C=ChestX-Ray14, CL=Clavicle, CM=Cardiomegaly, H=Heart, L=Lung, M=MIMIC-CXR, PM=Pneumonia, R=Rib, TB=Tuberculosis, Z=Other. \\ 
\textbf{Datasets}: C=ChestX-ray14, J=JSRT+SCR, M=MIMIC-CXR, MO=Montgomery, O=Open-i, PL=PLCO, PP=Ped-pneumonia, PR=Private, RP=RSNA-Pneumonia, S=Shenzen, SM=Simulated CXR from CT.}
\label{tab:domain_adaptation_table}
\begin{tabular}{p{3.6cm}p{8.75cm}p{1.2cm}p{1.35cm}p{1.35cm}}
\toprule
      Citation &                                                                          Method &       Other Tasks & Labels & Datasets \\
\midrule
\cite{Oliv20b} &                                      MUNIT based DA model for lung segmentation &       \textbf{SE} & CL,H,L &        J \\
 \cite{Dong18} &                      Adversarial training of lung and heart segmentation for DA &       \textbf{SE} &     CM &     J,PR \\
 \cite{Zhan18} & CycleGAN guided by a segmentation module to convert CXR to CT projection images &       \textbf{SE} &  H,L,Z &       PR \\
\cite{Chen18a} &          CycleGAN based DA model with semantic aware loss for lung segmentation &       \textbf{SE} &      L &       MO \\
\cite{Oliv20a} &                                 Conditional GANs based DA for bone segmentation &       \textbf{SE} &      R &       SM \\
 \cite{Leng20} &                    Continual learning methods to classify data from new domains &       \textbf{IL} &    C,M &      C,M \\
\cite{Tang19a} &     CycleGAN model to adapt adult to pediatric CXR for pneumonia classification &       \textbf{IL} &     PM &    PP,RP \\
 \cite{Maha19} &     Conditional GAN based DA for image registration using segmentation guidance & \textbf{IG},RE,SE &      L &        C \\
 \cite{Mada18} &    Adversarial based method adapting new domains for abnormality classification &    \textbf{IG},IL &     CM &       PL \\
 \cite{Zech18} &                  Assessment of generalization to data from different institutes &                IL &     PM &      C,O \\
 \cite{Sath20} &     Demonstrates the effect of training and test on data from different domains &                IL &     TB &        S \\
\bottomrule
\end{tabular}
\end{table*}

%% file: tables/Other.tex
\begin{table*}[h!]
\centering
\scriptsize
\caption{Other Studies (Section \ref{sec:studies_oth}).\\ 
\textbf{Tasks}: IL=Image-level Predictions, IR=Image Retrieval, OD=Out-of-Distribution, RE=Registration, RG=Report Generation, RP=Report Parsing. \textbf{Bold font} in tasks implies that this additional task is central to the work and the study also appears in another table in this paper.\\ 
\textbf{Labels}: C=ChestX-Ray14, H=Heart, L=Lung, Q=Image Quality, T=Triage/Abnormal, TB=Tuberculosis, X=CheXpert, Z=Other. \\ 
\textbf{Datasets}: C=ChestX-ray14, J=JSRT+SCR, M=MIMIC-CXR, MO=Montgomery, O=Open-i, PR=Private, S=Shenzen, X=CheXpert.}
\label{tab:other_table}
\begin{tabular}{p{3.6cm}p{8.75cm}p{1.2cm}p{1.35cm}p{1.35cm}}
\toprule
      Citation &                                                                           Method &          Tasks & Labels & Datasets \\
\midrule
 \cite{Owai20} &   Uses a database of the intermediate ResNet-50 features to find similar studies & \textbf{IL},IR &     TB &     MO,S \\
 \cite{Syed20} &  Generate reports by classifying CXRs, and finding and modifying similar reports &          RG,RP &      Z &      C,M \\
  \cite{Li20c} &   Extracts features from Chest X-rays and uses another network to write reports. &          RG,IL &      C &      C,O \\
 \cite{Yuan19} & Generates radiology reports by training on classification labels and report text &          RG,IL &      Z &      O,X \\
 \cite{Xue18b} &                    A novel recurrent generation network with attention mechanism &             RG &      Z &        O \\
 \cite{Mans20} &              Anatomical priors to improve deep learning based image registration &             RE &    H,L &   J,MO,S \\
 \cite{Marq19} &          Proposes a method to reject out-of-distribution images during test time &          OD,IL &      Z &        C \\
 \cite{Bozo20} &          Proposes to detect anomalies based on a dataset of autoencoder features &             OD &    Q,T &        C \\
\cite{Call19a} &     Mahalanobis distance on network layers to detect out-of-distribution samples &             OD &      Z &        C \\
 \cite{Anav15} &       Compares the extracted feature and classification similarities for ranking &             IR &        &       PR \\
  \cite{Haq21} &        Uses extracted features to cluster similarly labeled CXRs across datasets &             IR &    C,X &      C,X \\
 \cite{Chen18} &              Proposes a learnable hash to retrieve CXRs with similar pathologies &             IR &      Z &        C \\
 \cite{Conj17} &                   Residual network to retrieve images with similar abnormalities &             IR &      Z &        O \\
 \cite{Anav16} &           Combines features extracted from CXRs and metadata for image retrieval &             IR &      Z &       PR \\
 \cite{Silv20} &    Proposes to use the saliency maps as a similarity measure for image retrieval &             IR &      Z &        X \\
\bottomrule
\end{tabular}
\end{table*}

%% file: tables/commercial_table.tex
\begin{table*}[h!]
\centering
\scriptsize
\caption{Commercial Products for CXR analysis. (Section \ref{sec:commercial}) \\ \textbf{Labels}: T=Triage/Abnormal, PM=Pneumonia, CV=COVID, TB=Tuberculosis, LO=Lesion or Opacity, CM=Cardiomegaly, ND=Nodule, PE=Effusion, PT=Pneumothorax, TU=Catheter or Tube, LC=Lung Cancer, BS=Bone Suppression, E=Edema, Z=Other\\ \textbf{Output}: LOC=Localization, PRI=Prioritization, REP=Report, SCOR=Scoring}
\label{tab:commercial}
\begin{tabular}{p{2.3cm}p{3.4cm}p{4.9cm}p{2.9cm}p{2.6cm}}
\toprule
       Company & Product & Literature (4 most recent) & Labels (Total number) & Output\\
\midrule
Siemens Healthineers & AI-Rad Companion Chest X-Ray & ~\cite{Fisc20} & LO PE PT Z (5) &  LOC, SCOR, REP \\
Samsung Healthcare &	Auto Lung Nodule Detection & 	~\cite{Sim19} & ND (1) & LOC\\
Thirona	& CAD4COVID-XRay	& ~\cite{Murp20} &  CV (1) & LOC, SCOR\\
Thirona & 	CAD4TB &	~\cite{Murp20a, Habi20, Qin19,  Sant20} & TB (1) & LOC, SCOR\\
Oxipit &	ChestEye CAD & & T (1)  & REP (healthy) \\	
Arterys & 	Chest $|$ MSK AI	& & LO, ND, PE, PT (4) & LOC, SCOR, PRI\\
Quibim	& Chest X-Ray Classifier	&~\cite{Lian20} &  PM  CM  ND  PE  PT  E  Z (16) & LOC, SCOR, REP\\ 
GE	& Critical Care Suite  & & PT (1) & LOC, SCOR\\	
InferVision	& InferRead DR Chest & &  TB  PE  PT  LC  Z (9)  & LOC, SCOR\\	
JLK	& JLD-O2K  & &  LC  Z (16) & LOC, SCOR\\	
Lunit &	Lunit INSIGHT CXR	&~\cite{Hwan19, Hwan19a, Hwan19b, Qin19}&  TB  CM  ND  PE  PT  Z (11) & LOC, SCOR, PRI, REP\\
qure.ai	& qXR	& ~\cite{Sing18, Nash20, Engl20, Qin19} &  T  CV  TB  Z (30) & LOC, SCOR, PRI, REP\\
Digitec	& TIRESYA & &  BS (1) & Bone Suppressed Image\\	
VUNO &	VUNO Med-Chest X-Ray& 	~\cite{Kim17} &  LO  ND  PE  PT  Z (5) & LOC, SCOR\\
Riverain Technologies	& ClearRead Xray - Bone Suppress &	~\cite{Homa20, Dell17, Scha16, Scha14}& BS(1) & Bone Suppressed Image\\
Riverain Technologies	& ClearRead Xray - Compare	& & LC(1) & Subtraction Image\\
Riverain Technologies	& ClearRead Xray - Confirm	& & TU(1) & LOC\\
Riverain Technologies	& ClearRead Xray - Detect	& ~\cite{Dell17, Scha14a, Szuc13}& ND LC (2) & LOC\\
behold.ai	& Red Dot	& & T PT (2) & LOC\\
Zebra Medical Vision &	Triage Pleural Effusion	& & PE & LOC, PRI\\
Zebra Medical Vision &	Triage Pneumothorax	& & PT & LOC, PRI\\
\bottomrule
\end{tabular}
\end{table*}


